\documentclass[twocolumn,times]{aastex63}
\usepackage{threeparttable}

\usepackage{txfonts}

\newcommand{\dsct}{$\delta$~Scuti }

\begin{document}

\title{TESS Observations of Seven Newly Identified High-Amplitude \dsct Stars}
\shorttitle{TESS Observations of Seven New Identified HADS}
\shortauthors{Lv, C. et al.}

\author{Chenglong Lv \href{https://orcid.org/0000-0001-6354-1646}{\includegraphics[scale=0.4]{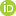}}}
\affil{Xinjiang Astronomical Observatory, Chinese Academy of Sciences, Urumqi, Xinjiang 830011, People's Republic of China}
\affil{School of Astronomy and Space Science, University of Chinese Academy of Sciences, Beijing 100049, People's Republic of China}

\author{Ali Esamdin \href{https://orcid.org/0000-0003-1845-4900}{\includegraphics[scale=0.4]{orcid.png}}}
\email{aliyi@xao.ac.cn}
\affil{Xinjiang Astronomical Observatory, Chinese Academy of Sciences, Urumqi, Xinjiang 830011, People's Republic of China}
\affil{School of Astronomy and Space Science, University of Chinese Academy of Sciences, Beijing 100049, People's Republic of China}

\author{A. Hasanzadeh \href{https://orcid.org/0000-0002-7286-1438}{\includegraphics[scale=0.4]{orcid.png}}}
\affil{Centre for Fusion, Space and Astrophysics, Department of Physics, University of Warwick, Coventry CV4 7AL, UK}
\affil{Institute of Geophysics, University of Tehran, Tehran, Iran}

\author{Shuguo Ma \href{https://orcid.org/0000-0001-5066-5682}
{\includegraphics[scale=0.4]{orcid.png}}}
\affil{Key Laboratory of Optical Astronomy, National Astronomical Observatories, Chinese Academy of Sciences, Beijing, 100012, People’s Republic of China}

\author{TaoZhi Yang \href{https://orcid.org/0000-0002-1859-4949}
{\includegraphics[scale=0.4]{orcid.png}}}
\affil{Ministry of Education Key Laboratory for Nonequilibrium Synthesis and Modulation of Condensed Matter, School of Physics, Xi’an Jiaotong University, Xi’an 710049, People’s Republic of China}

\author{Rivkat Karimov}
\affil{Ulugh Beg Astronomical Institute, Uzbekistan Academy of Sciences, Tashkent 100052, Uzbekistan}

\journalinfo{The Astrophysical Journal}

\begin{abstract}

We report seven newly identified High-amplitude \dsct (HADS) stars. Among them, two stars (TIC 30977864, TIC 387379145) exhibit pure radial pulsation without the excitation of non-radial modes. TIC 30977864 is classified as a double-mode HADS star, while the other four stars potentially show triple-mode HADS behavior. TIC 112682462 and TIC 255603395 closely resemble RR Lyrae stars based on their light curves, position in the period-luminosity diagram, and the period-ratio diagram. However, without spectral observations, it is challenging to ascertain whether these two stars are HADS or RR Lyrae stars. TIC 281695001 exhibits a fundamental frequency amplitude smaller than that of the first overtone, suggesting the presence of intriguing phenomena that necessitate further analysis. We analyzed the metallicities and period by using \citet{Netzel2022} derived from the model. The metallicities of the 176 stars display a broad distribution ranging from -2.0 dex to 0.5 dex, with periods spanning 0.05 to 0.20 days. This random distribution of metallicities may contribute to the dispersion observed in the P1/P0 ratio. To derive more accurate conclusions, future spectroscopic observations of a larger sample of HADS stars are crucial. These observations will provide precise rotational velocities and more accurate determinations of metallicities.

\end{abstract}

\keywords{asteroseismology -- stars: oscillations -- stars: variables: \dsct -- stars: variables: HADS}

\section{Introduction}

In recent years, significant advancements have been made in the development of space telescopes. Noteworthy examples include the Microvariability and Oscillations of STars (MOST) telescope \citep{2003PASP..115.1023W}, the Convection, Rotation, and Planetary Transits (CoRoT) mission \citep{2009A&A...506..411A}, and the Kepler mission \citep{2010PASP..122..131G,2010ApJ...713L..79K}. These telescopes have revolutionized the field of asteroseismology by providing a wealth of high-resolution data, facilitating in-depth analyses of pulsating stars' physical structure \citep{1994ARA&A..32...37B,Dupret2004,ASTERO,2015pust.book.....C,2019LRSP...16....4G,Daszy2022}. Among these missions, the Transiting Exoplanet Survey Satellite (TESS) has made a particularly notable contribution. TESS, during its two-year primary mission, has extensively surveyed nearly 85\% of the sky across 26 sectors \citep{2016ApJ...830..138C,2019AJ....157..245H}, providing a vast database of high-quality observational data. The utilization of this comprehensive dataset for research purposes will have a profound impact on the advancement of asteroseismology and our understanding of stellar structure and evolution.

High-amplitude \dsct (HADS) stars form a specific subgroup within the broader class of \dsct stars. These stars are characterized by their slow rotation, typically exhibiting $v$ sin $i$ values below 30 km s$^{-1}$, and their pulsation periods falling within the range of 1 to 6 hours. Furthermore, HADS stars display peak-to-peak light amplitudes exceeding 0.3 mag \citep{McNamara2000}. Historically, prior to the availability of precise space-based data, observations of HADS stars often revealed the presence of one or two radial pulsation modes, primarily associated with the fundamental and/or first overtone mode \citep{Poretti2011,Bowman2021,lv2022AJ,Yang2022}. However, recent advancements in sophisticated data processing techniques, as demonstrated by \citet{Lares2020}, have enabled the identification of low-amplitude frequencies in the frequency spectrum of HADS stars as well \citep{Poretti2011,Bowman2021}. By analyzing the frequency ratios, it becomes possible to determine the radial modes present in HADS stars \citep{Petersen1973}. Subsequently, by constructing models based on these radial frequencies and comparing them with observed ratios, various fundamental parameters of the star can be derived, including its evolutionary stage \citep{Bowman2021,Daszy2022,lv2022ApJ}. For example, \citet{Daszy2020} conducted detailed seismic modeling of SX Phe, using high-precision photometry from the TESS mission, confirming its post-main sequence evolutionary stage \citep{Daszy2020}. In a study by \citet{Bowman2017}, the results of a search for amplitude modulation of pulsation modes in 983 \dsct stars were presented, with only two HADS stars included in their sample.

\citet{Mow2016} conducted observations and noted that the pulsation frequencies of HADS stars did not exhibit significant changes. However, they made an intriguing observation regarding the amplitudes of different pulsation modes. While the amplitudes of the fundamental and first overtone modes remained consistent across the two datasets, the amplitude of the second overtone mode increased by 44\%, which corresponds to an almost 12 standard deviation change. This finding represents a previously unreported type of evolution in HADS stars. The significant and rapid amplitude change observed in the second overtone mode of GSC 03144-595 suggests that this mode may exhibit transient behavior. The transient nature of the second overtone mode could potentially explain the rarity of triple-mode pulsators in HADS stars. If the second overtone modes appear and disappear over short timescales, their observation would require specific timing observations of the star within a particular window. Consequently, the discovery and detailed study of more HADS stars are crucial for enhancing our comprehension of this specific type of pulsating star.

In this paper, Section 2 provides an introduction to the TESS photometry data and the Fourier analysis method used in the study. Section 3 presents the results of the frequency analysis conducted on seven HADS stars. The analysis includes the identification and characterization of pulsation frequencies in each star. In Section 4, the distribution of the period ratios for the HADS stars is examined and analyzed. Finally, Section 5 offers a concise summary of the main findings and conclusions derived from the study.

\section{TESS photometry and Fourier analysis}

We primarily utilized photometric data processed by the TESS Science Processing Operations Center (SPOC) \citep{Jenkins2016}, which can be accessed through the Mikulski Archive for Space Telescopes (MAST)\footnote{https://archive.stsci.edu/}. After carefully screening the catalog of more than 59,000 \dsct stars provided by \citet{Zhou2023}, we identified seven HADS stars through SPOC photometric data. To complement our analysis, we cross-matched our sample with the LAMOST DR9 catalog \citep{Cui2012}\footnote{http://lamost.org/dr9/}, but only obtained stellar parameters for TIC 353050847.

The SPOC photometric data consists of Simple Aperture Photometry (SAP) and Pre-Search Data Conditioning (PDC) SAP flux, with the latter specifically designed to eliminate long-term trends \citep{Twicken2010}. In our data processing, we focused on the PDC SAP flux, which not only corrects for onboard systematics but also considers the contributions of neighboring stars to the overall flux. Additionally, to ensure the quality of the data, we applied a 4.5$\sigma$ flux clipping procedure to remove any significant outliers. To further refine the dataset, the average value for each quarter is then subtracted to obtain the corrected flux, then the data has been corrected for systematic errors such as the cooling down, warming up, outliers, and jumps. Finally, we calculated the amplitude of the flux variations, which serves as a crucial parameter for our analysis.

For 2-minute cadence observations, the Nyquist frequency, denoted as $f_N$, is calculated to be $f_N = 360 \text{ day}^{-1}$, which comfortably exceeds the frequency range of 5 to 80 day$^{-1}$ from which we extracted frequencies. This frequency range covers the typical pulsation interval observed in \dsct stars. To distinguish closely spaced frequencies in the power spectrum, we employ the resolution frequency $f_\mathrm{res} = 1.5 / \Delta T$ \citep{Loumos1978}, where $\Delta T$ represents the duration of the dataset. If the difference between two frequencies is larger than the resolution frequency, we consider those frequencies to be resolved. To identify significant frequencies in the amplitude spectra, we utilized the FELIX tool \citep{Charpinet2010,Zong2016}, and then fitted the rectified light curve using the following formula:
\begin{equation}
m = m_0 + \sum_{i=1}^N A_i \sin(2\pi(f_i t + \phi_i)), \label{equation1}
\end{equation}
In our analysis, we assign $m_0$ as the zero-point, $A_i$ as the amplitude, $f_i$ as the frequency, and $\phi_i$ as the corresponding phase. To identify significant frequencies, we typically consider the highest peaks in the power spectrum as significant frequencies candidates. Subsequently, we perform a multi-frequency least square fit of the light curve, utilizing Equation \ref{equation1}, for all the detected significant frequencies. This allows us to obtain solutions for each frequency by minimizing the residuals. The residuals are obtained by subtracting the theoretical light curve, constructed using the obtained solutions, from the rectified data. We then search for additional frequencies by analyzing the residuals and repeating the steps mentioned above until no significant peaks are found in the spectrum. We refine the set of previously identified frequency values whenever new frequencies are detected. To determine the significance of a detected peak, we employ a signal-to-noise ratio (S/N) criterion with a threshold of 5.2, following the recommendation of \citet{Baran2015}. The uncertainty in frequency is determined using the method proposed by \citet{1999DSSN...13...28M}, allowing us to estimate the accuracy of our frequency measurements.

Mode identification of the extracted significant frequencies in the frequency analysis is extremely important, and correctly identifying the mode of the frequencies facilitates the modeling of the target and thus improves our understanding and knowledge of the class of stars. When dealing with data sources like $Kepler$ or TESS, which provide single-band observations, standard techniques such as rotational splitting, frequency spacing, and echelle diagrams \citep{ASTERO} are commonly employed to determine the spherical degree (l) and azimuthal order (m). These methods have been extensively utilized in numerous studies found in the literature. For instance, \citet{Tian2014,Kern2018} utilized single-band $Kepler$ data to determine the spherical degree (l) using period spacing and rotational splitting. With the availability of high-quality continuous space-based data from missions like $Kepler$ and TESS, recent asteroseismological investigations have primarily focused on frequency splitting and frequency spacing methods for determining (l) and (m). Since all of our targets in this study are HADS stars, and most of the frequencies are radial and combinations of these radial frequencies, trying to accurately identify modes of non-radial frequencies by the method described above is very difficult.

\section{Observations and analysis of the newly identified HADS}
\label{sect:HADS}

\paragraph{TIC 30977864}
(V* V494 Sgr; $\alpha_{2000}$=19$^{h}$:11$^{m}$:14.451$^{s}$, $\delta_{2000}$=-34$\degr$:53$\arcmin $:46.068$\arcsec$) is a previously known \dsct star confirmed by \citet{Guthnick1933}. In our analysis, we have identified this star as a new double-mode HADS star. The TESS Space Telescope observed TIC 30977864 during Sector 27 for a duration of 23.5 days, spanning from Barycentric Julian Date (BJD) 2459036.278881 to 2459059.779851, with a cadence of 2 minutes. In our study, we utilize the corrected flux from each point and then convert it to magnitude. The same data processing method is applied to the remaining stars in our sample.

\begin{figure}[htp!]
\begin{center}
  \includegraphics[width=0.48\textwidth]{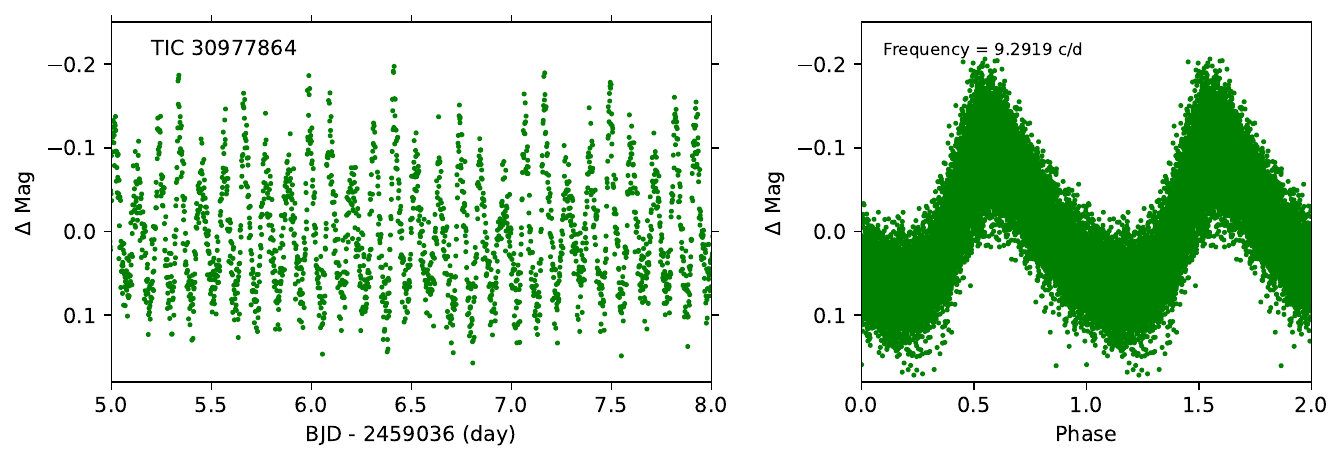}
  \includegraphics[width=0.482\textwidth]{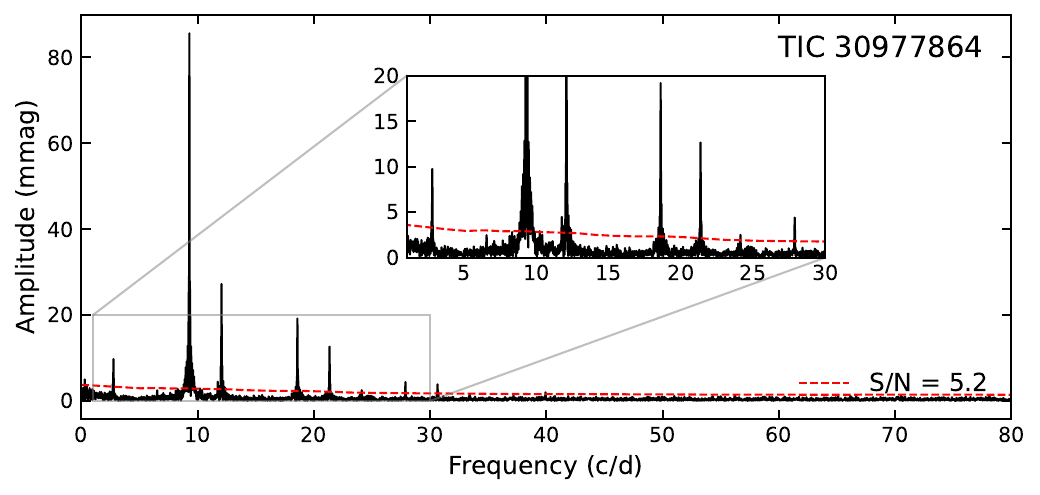}
  \caption{The upper left panel shows a portion of the light curve of TIC 309777864 for 3 days. The upper right panel shows the phase diagram and the lower panel shows Fourier amplitude spectra.}
    \label{fig:TIC 30977864_light_fre}
\end{center}
\end{figure}

Following the data processing steps described above, we obtained a rectified light curve consisting of 15,833 data points. The upper left panel of Figure \ref{fig:TIC 30977864_light_fre} presents a portion of the rectified light curve for this star covering 3 days. From this figure, it is evident that the peak-to-peak amplitude of this star is approximately 0.32 mag, which is consistent with the typical amplitudes observed in HADS stars.

Through Fourier analysis of TIC 30977864, we detected a total of 11 frequencies. Theoretical models, as described by \citet{Stellingwerf1979}, predict period ratios for the first two radial modes as $f_{1}$ / $f_{2}$ = (0.756 - 0.787). In the case of TIC 30977864, the frequency ratio of 0.770696 between the two high-amplitude independent modes confirms their identification as the fundamental and first overtone radial modes \citep{Breger2000}. The upper right panel of Figure \ref{fig:TIC 30977864_light_fre} shows the phase diagram of TIC 30977864, folded by the fundamental frequency F0 = 9.2919(1)~day$^{-1}$. Furthermore, the lower panel of Figure \ref{fig:TIC 30977864_light_fre} presents the Fourier amplitude spectra. To identify combination frequencies, we searched for linear sum and difference frequencies, $n$$\nu$$_{i}$ $\pm$ $m$$\nu$$_{j}$, using the tolerance criterion proposed by \citet{Loumos1978}. We assumed that the highest-amplitude peaks within a combination family correspond to the real pulsation mode frequencies \citep{Kurtz2015}. This method was applied to each of the remaining stars to identify combination frequencies. Thus, Table \ref{Tab30977864} lists the fundamental frequency $f_{1}$ (labeled 'F0'), first overtone $f_{2}$ (labeled 'F1'), combination frequencies ($f_{4}$, $f_{5}$, $f_{7}$...$f_{11}$), and harmonic frequencies ($f_{3}$, $f_{6}$, $f_{9}$) of the two radial modes. Additionally, we detected a non-radial frequency ($f_{8}$) for which frequency combination was not possible. Based on the frequency values, we assume that this frequency should be a low-order non-radial p mode.

\begin{deluxetable}{ccccccc}
\renewcommand\arraystretch{1.2}
\tabletypesize{\scriptsize}
\setlength\tabcolsep{3.5pt}
\tablewidth{\textwidth}
\tablenum{1}
\tablecaption{The Pulsation Mode Frequencies in SC data of TIC 30977864. \label{Tab30977864}}
\tablehead{
\colhead{$f_{i}$}               &
\colhead{Frequency (day$^{-1}$)}  &
\colhead{Amplitude (mmag)}      &
\colhead{Phase (rad)}           &
\colhead{S/N}                   &
\colhead{Comment}               &
}
\startdata
1	&	9.2919 	$\pm$	0.0001 	&	85.8 	$\pm$	0.4 	&	0.915	$\pm$	0.002	&	194.7	&F0		\\
2	&	12.0565 $\pm$	0.0004 	&	27.1 	$\pm$	0.4 	&	0.999	$\pm$	0.005	&	62.5	&F1		\\
3	&	18.5843 $\pm$	0.0005 	&	19.2 	$\pm$	0.4 	&	0.953	$\pm$	0.006	&	50.2	&2F0		\\
4	&	21.3485 $\pm$	0.0007 	&	12.8 	$\pm$	0.4 	&	0.05	$\pm$	0.01	&	33.7	&F0+F1		\\
5	&	2.763 	$\pm$	0.001 	&	9.6 	$\pm$	0.4 	&	0.11	$\pm$	0.02	&	19.8	&F1-F0		\\
6	&	27.877 	$\pm$	0.002 	&	4.4 	$\pm$	0.4 	&	0.001	$\pm$	0.03	&	11.7	&3F0	\\
7	&	30.639 	$\pm$	0.002 	&	3.9 	$\pm$	0.4 	&	0.12	$\pm$	0.03	&	10.6	&2F0+F1		\\
8	&	11.741 	$\pm$	0.003 	&	3.3 	$\pm$	0.4 	&	0.08	$\pm$	0.04	&	7.5	    &-		\\
9	&	24.116 	$\pm$	0.003 	&	2.6 	$\pm$	0.4 	&	0.13	$\pm$	0.05	&	7.0	    &2F1 		\\
10	&	6.529 	$\pm$	0.005 	&	2.3 	$\pm$	0.5 	&	0.98	$\pm$	0.06	&	5.1	    &2F0-F1		\\
11	&	39.931 	$\pm$	0.005 	&	1.9 	$\pm$	0.4 	&	0.16	$\pm$	0.06	&	5.1	    &3F0+F1		\\
   \enddata
    \tablecomments{Among these frequencies, two peaks are independent frequencies, others are harmonic or combinations (denoted by $f_{i}$). '-' represents non-radial frequencies.}
\end{deluxetable}

\paragraph{TIC 90322352}

\begin{figure}[htp!]
\begin{center}
  \includegraphics[width=0.48\textwidth]{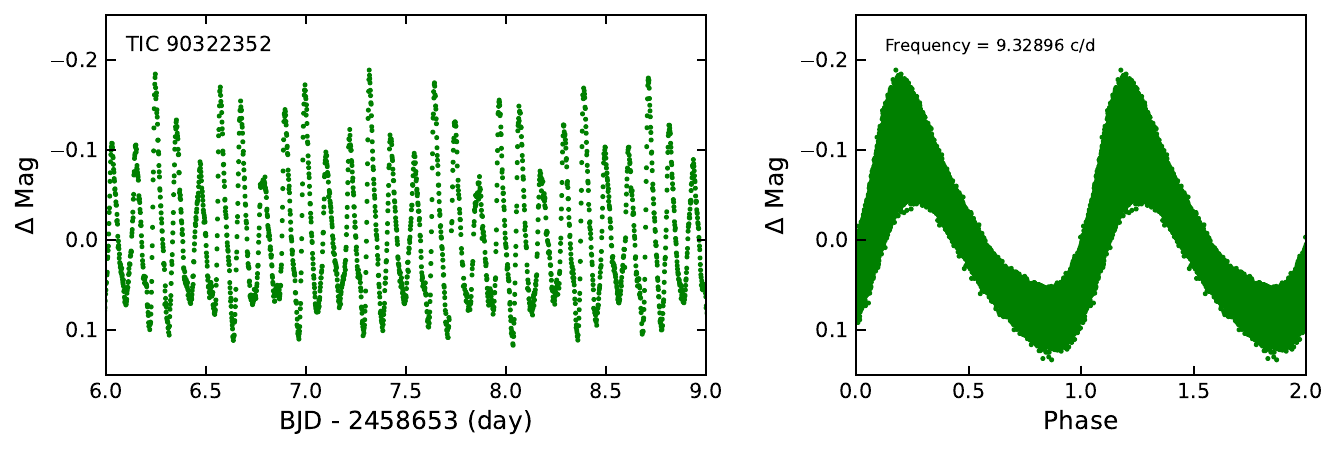}
  \includegraphics[width=0.482\textwidth]{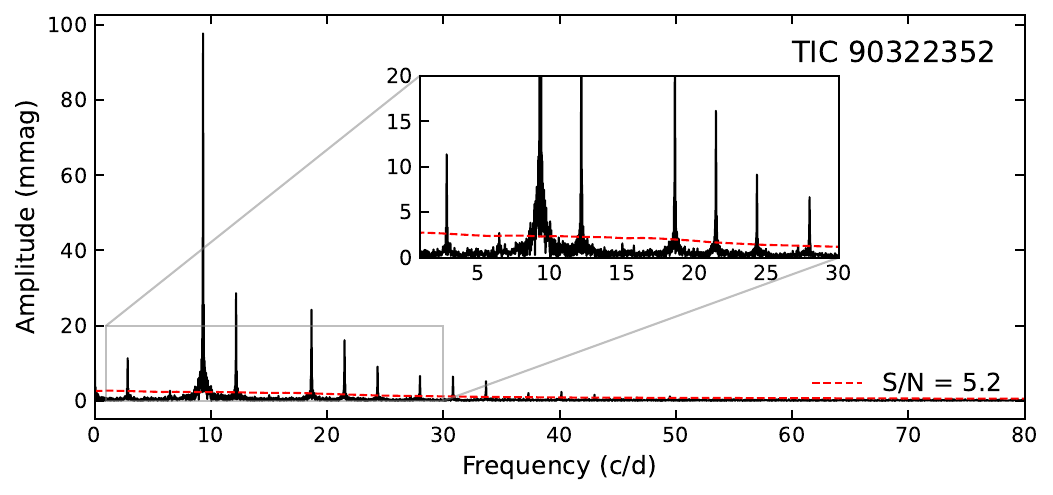}
  \caption{The upper left panel shows a portion of the light curve of TIC 90322352. The upper right panel shows the phase diagram and the lower panel shows Fourier amplitude spectra. The phase diagram shows that the light variation has a tendency to climb rapidly and fall slowly, which is typical of HADS stars.}
    \label{fig:TIC 90322352_light_fre}
\end{center}
\end{figure}

(TYC 7909-809-1; $\alpha_{2000}$=18$^{h}$:25$^{m}$:36.252$^{s}$, $\delta_{2000}$=-42$\degr$:13$\arcmin $:35.847$\arcsec$) is a known \dsct star identified by \citet{Barac2022}. Through the Fourier analysis, we suggest that this star might be a new triple-mode HADS star. TESS observed TIC 90322352 during Sector 13 for a duration of 28.4 days, from BJD 2458653.920490 to 2458682.361868. After processing, we obtained a rectified light curve consisting of 17392 data points. The upper left panel of Figure \ref{fig:TIC 90322352_light_fre} shows a segment of the rectified light curve. From the figure, we observe a peak-to-peak amplitude of approximately $\sim$0.30 mag for this star.

We detected a total of 26 frequencies through the Fourier analysis. The frequency ratio of the two high-amplitude independent modes for TIC 90322352 is 0.766471, indicating that they correspond to the fundamental and first overtone radial modes \citep{Breger2000}. Table \ref{Tab90322352} lists the fundamental frequency $f_{1}$ as 'F0', the first overtone $f_{2}$ as 'F1', and the combination frequencies (i.e., $f_{4}$, $f_{5}$, $f_{8}$...$f_{22}$) and harmonics frequencies (i.e., $f_{3}$, $f_{6}$, $f_{7}$...$f_{19}$) associated with these two radial modes.

\citet{Stellingwerf1979} conducted a statistical analysis of pulsation modes in several \dsct stars and presented the period ratios of the second radial mode as $f_{0}$ / $f_{2}$ = (0.611 - 0.632). For TIC 90322352, we measured the ratio $f_{0}$ / $f_{2}$ as 0.621425, but due to the small amplitude of F2, we could not precisely determine it as a second overtone frequency. Further certification is needed regarding the mode of this frequency. The upper right panel of Figure \ref{fig:TIC 90322352_light_fre} shows the phase diagram of TIC 90322352. The lower panel of Figure \ref{fig:TIC 90322352_light_fre} presents the amplitude spectra. No non-radial frequencies other than radial were detected for this target.

\begin{deluxetable}{ccccccc}
\renewcommand\arraystretch{1.2}
\tabletypesize{\scriptsize}
\setlength\tabcolsep{1.5pt}
\tablewidth{0.5\textwidth}
\tablenum{2}
\tablecaption{The Pulsation Mode Frequencies in SC data of TIC 90322352. \label{Tab90322352}}
\tablehead{
\colhead{$f_{i}$}               &
\colhead{Frequency (day$^{-1}$)}  &
\colhead{Amplitude (mmag)}      &
\colhead{Phase (rad)}           &
\colhead{S/N}                   &
\colhead{Comment}               &
}
\startdata
1	&	9.32896 	$\pm$	0.00002 &	98.58 	$\pm$	0.08 	&	0.2456 	$\pm$	0.0002 	&	1304.7 	&F0	\\
2	&	12.17132 	$\pm$	0.00005 &	28.86 	$\pm$	0.07 	&	0.0482 	$\pm$	0.0008 	&	397.1 	&F1	\\
3	&	18.65784 	$\pm$	0.00006 &	24.64 	$\pm$	0.06 	&	0.6418 	$\pm$	0.0008 	&	388.5 	&2F0	\\
4	&	21.50010 	$\pm$	0.00008 &	16.80 	$\pm$	0.06 	&	0.488 	$\pm$	0.001 	&	270.8 	&F0+F1	\\
5	&	2.8424   	$\pm$	0.0001 	&	11.82 	$\pm$	0.08 	&	0.817 	$\pm$	0.002 	&	143.2 	&F1-F0	\\
6	&	24.3423 	$\pm$	0.0001 	&	8.99 	$\pm$	0.06 	&	0.455 	$\pm$	0.002 	&	148.0 	&2F1	\\
7	&	27.9866 	$\pm$	0.0001 	&	6.98 	$\pm$	0.06 	&	0.040 	$\pm$	0.002 	&	118.6 	&3F0	\\
8	&	30.8291 	$\pm$	0.0001 	&	6.53 	$\pm$	0.06 	&	0.896 	$\pm$	0.002 	&	111.7 	&2F0+F1	\\
9	&	33.6715 	$\pm$	0.0002 	&	4.95 	$\pm$	0.06 	&	0.054 	$\pm$	0.003 	&	86.6 	&F0+2F1	\\
10	&	6.4863 	    $\pm$	0.0005 	&	3.08 	$\pm$	0.08 	&	0.554 	$\pm$	0.008 	&	39.2 	&2F0-F1	\\
11	&	40.1579  	$\pm$	0.0004 	&	2.58 	$\pm$	0.06 	&	0.151 	$\pm$	0.006 	&	46.8 	&3F0+F1	\\
12	&	37.3158  	$\pm$	0.0005 	&	2.05 	$\pm$	0.06 	&	0.437 	$\pm$	0.008 	&	36.3 	&4F0	\\
13	&	15.0122  	$\pm$	0.0009 	&	1.61 	$\pm$	0.07 	&	0.90 	$\pm$	0.01 	&	23.8 	&F2	\\
14	&	43.0001 	$\pm$	0.0008 	&	1.48 	$\pm$	0.05 	&	0.29 	$\pm$	0.01 	&	27.1 	&2F0+2F1	\\
15	&	49.4861  	$\pm$	0.0009 	&	1.23 	$\pm$	0.05 	&	0.55 	$\pm$	0.01 	&	23.0 	&4F0+F1	\\
16	&	36.513     	$\pm$	0.001 	&	1.19 	$\pm$	0.06 	&	0.44 	$\pm$	0.01 	&	21.3 	&3F1	\\
17	&	15.815     	$\pm$	0.001 	&	1.15 	$\pm$	0.07 	&	0.98 	$\pm$	0.01 	&	17.3 	&3F0-F1	\\
18	&	27.186     	$\pm$	0.001 	&	1.10 	$\pm$	0.06 	&	0.22 	$\pm$	0.01 	&	18.6 	&F1+F2	\\
19	&	46.644     	$\pm$	0.001 	&	0.79 	$\pm$	0.05 	&	0.86 	$\pm$	0.02 	&	14.7 	&5F0	\\
20	&	5.687 	    $\pm$	0.002 	&	0.59 	$\pm$	0.08 	&	0.65 	$\pm$	0.04 	&	7.4 	&F1-F2	\\
21	&	45.845     	$\pm$	0.002 	&	0.56 	$\pm$	0.05 	&	0.08 	$\pm$	0.03 	&	10.5 	&F0+3F1	\\
22	&	25.144     	$\pm$	0.003 	&	0.40 	$\pm$	0.06 	&	0.41 	$\pm$	0.04 	&	6.7 	&4F0-F1	\\
   \enddata
    \tablecomments{Among these frequencies, two peaks are independent frequencies, others are harmonic or combinations (denoted by $f_{i}$).}
\end{deluxetable}

\paragraph{TIC 112682462}

\begin{figure}[htp!]
\begin{center}
  \includegraphics[width=0.48\textwidth]{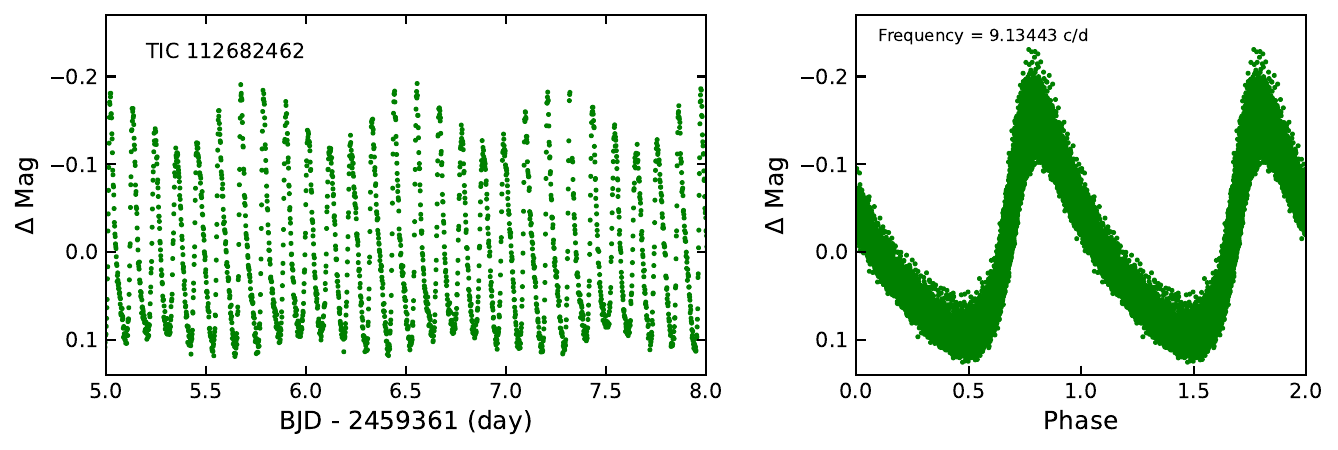}
  \includegraphics[width=0.482\textwidth]{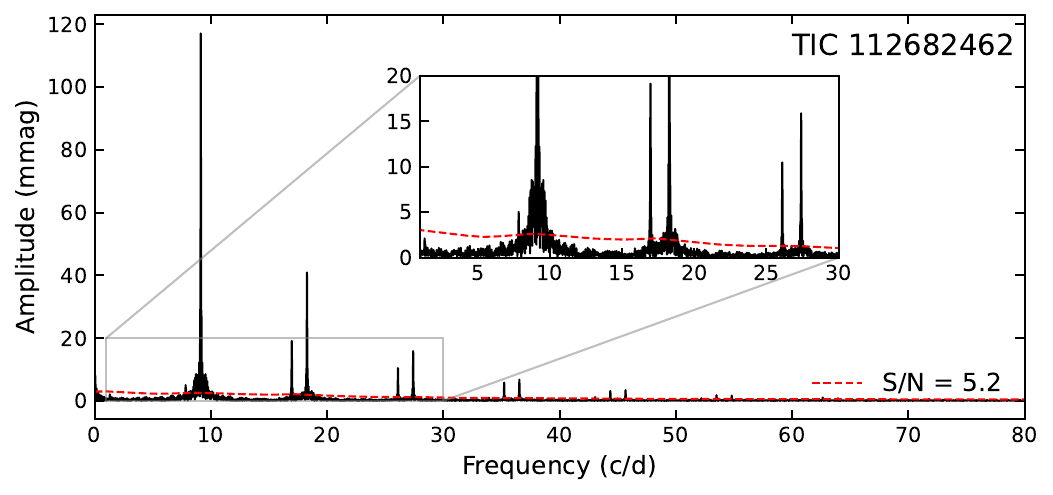}
  \caption{The upper left panel shows a portion of the light curve of TIC 112682462. The upper right panel shows the phase diagram folded by the fundamental frequency F0 = 9.13443(2)~day$^{-1}$ and the lower panel shows Fourier amplitude spectra.}
    \label{fig:TIC 112682462_light_fre}
\end{center}
\end{figure}

(V* V557 Sco; $\alpha_{2000}$=17$^{h}$:58$^{m}$:24.543$^{s}$, $\delta_{2000}$=-41$\degr$:50$\arcmin $:30.425$\arcsec$) is a known \dsct star confirmed by \citet{Kotov1987}. We suggest that this star shows promising indications of being a new double-mode HADS star. TESS observed TIC 112682462 during Sector 39 for 27.7 days, with data collected at a 2-minute cadence from BJD 2459361.771158 to 2459389.525794, a total of 18217 points were obtained. The upper left panel of Figure \ref{fig:TIC 112682462_light_fre} shows a portion of the light curve, revealing a peak-to-peak amplitude of approximately 0.32 mag for the star. The light curve of TIC 112682462 exhibits a remarkable similarity to that of an RR Lyrae star \citep{Clementini2023}. Upon comparing the distribution of TIC 112682462 and TIC 255603395 with several other HADS stars in Figure \ref{fig:HR_HADS} and Figure \ref{fig:PLR}, it becomes evident that the positions of these two stars deviate from those of other HADS stars. However, without precise spectral parameters to substantiate our conclusion, we can only tentatively propose that these two stars may potentially be classified as RR Lyrae stars.

By analyzing the frequency spectrum of TIC 112682462, we detected a total of 28 frequencies. \citet{Stellingwerf1979} conducted a statistical analysis of pulsation modes in various \dsct stars and presented the period ratios of the third radial mode as $f_{0}$ / $f_{3}$ = (0.500 - 0.525), where $f_{3}$ represents the third overtone. The two high-amplitude independent modes observed in TIC 112682462 exhibit a frequency ratio of 0.538564, suggesting the possibility that they are the fundamental and third overtone radial modes. The upper right panel of Figure \ref{fig:TIC 112682462_light_fre} shows the phase diagram of TIC 112682462, folded by the fundamental frequency F0 = 9.13443(2)~day$^{-1}$. The lower panel of Figure \ref{fig:TIC 112682462_light_fre} shows the amplitude spectra for the 2-minute cadence data. Therefore, we list the fundamental frequency $f_{1}$ as 'F0', the third overtone frequency $f_{3}$ as 'F3', as well as the combination frequencies (i.e., $f_{5}$, $f_{7}$, $f_{10}$...$f_{28}$) and harmonic frequencies (i.e., $f_{2}$, $f_{4}$, $f_{6}$...$f_{26}$) of the two radial modes in Table \ref{Tab112682462}. The measured ratio of $f_{0}$ / $f_{3}$ is 0.538564, slightly larger than the theoretical ratio. However, since we detect numerous combined frequencies between these two independent frequencies, along with the significant amplitude value of F3, we assume that this star is a double-mode HADS star.

It is worth noting that for some reason, the first and second overtone frequencies are not detected. Additionally, we detected some frequencies in the low-frequency region that may correspond to g modes. These g modes might have a higher order depending on the frequency value. For these 4 g mode frequencies, we find that they have essentially the same frequency spacing, about 0.068~day$^{-1}$, which is larger than the Riley resolution of 0.054~day$^{-1}$. So it is extremely possible that these 4 frequencies are a quintuplet centered on $f_{20}$, that the 0.068~day$^{-1}$ might be the rotational splitting frequency, and that quintuplet could characterize the internal information of this target.

\begin{deluxetable}{ccccccc}
\renewcommand\arraystretch{1.2}
\tabletypesize{\scriptsize}
\setlength\tabcolsep{1.5pt}
\tablewidth{0.5\textwidth}
\tablenum{3}
\tablecaption{The Pulsation Mode Frequencies in SC data of TIC 112682462. \label{Tab112682462}}
\tablehead{
\colhead{$f_{i}$}               &
\colhead{Frequency (day$^{-1}$)}  &
\colhead{Amplitude (mmag)}      &
\colhead{Phase (rad)}           &
\colhead{S/N}                   &
\colhead{Comment}               &
}
\startdata
1	&	9.13443 	$\pm$	0.00002 &	117.3 	$\pm$	0.1 	&	0.1059 	$\pm$	0.0003 	&	1009.9 	&F0	\\
2	&	18.26891 	$\pm$	0.00004 &	41.3 	$\pm$	0.1 	&	0.3604 	$\pm$	0.0007 	&	448.2 	&2F0	\\
3	&	16.9607 	$\pm$	0.0001 	&	19.0 	$\pm$	0.1 	&	0.871 	$\pm$	0.002 	&	201.0 	&F3	\\
4	&	27.4032 	$\pm$	0.0001 	&	16.0 	$\pm$	0.1 	&	0.627 	$\pm$	0.002 	&	181.1 	&3F0	\\
5	&	26.0947 	$\pm$	0.0002 	&	10.4 	$\pm$	0.1 	&	0.093 	$\pm$	0.003 	&	119.0 	&F0+F3	\\
6	&	36.5381 	$\pm$	0.0002 	&	6.7 	$\pm$	0.1 	&	0.894 	$\pm$	0.004 	&	81.9 	&4F0	\\
7	&	35.2294 	$\pm$	0.0003 	&	6.0 	$\pm$	0.1 	&	0.333 	$\pm$	0.004 	&	72.3 	&2F0+F3	\\
8	&	7.8254 	    $\pm$	0.0005 	&	5.2 	$\pm$	0.1 	&	0.608 	$\pm$	0.007 	&	44.2 	&F3-F0	\\
9	&	45.6719 	$\pm$	0.0005 	&	3.3 	$\pm$	0.1 	&	0.176 	$\pm$	0.008 	&	41.5 	&5F0	\\
10	&	44.3641 	$\pm$	0.0005 	&	3.2 	$\pm$	0.1 	&	0.586 	$\pm$	0.008 	&	38.7 	&3F0+F3	\\
11	&	53.4983 	$\pm$	0.0009 	&	1.8 	$\pm$	0.1 	&	0.85 	$\pm$	0.01 	&	22.5 	&4F0+F3	\\
12	&	54.806 	$\pm$	0.001 	&	1.6 	$\pm$	0.1 	&	0.45	$\pm$	0.02 	&	19.6 	&6F0	\\
13	&	43.052 	$\pm$	0.001 	&	1.2 	$\pm$	0.1 	&	0.11 	$\pm$	0.02 	&	14.9 	&F0+2F3	\\
14	&	33.917 	$\pm$	0.001 	&	1.2 	$\pm$	0.1 	&	0.86 	$\pm$	0.02 	&	14.2 	&2F3	\\
15	&	10.443 	$\pm$	0.001 	&	1.6 	$\pm$	0.1 	&	0.03 	$\pm$	0.02 	&	13.9 	&2F0-F3	\\
16	&	62.632 	$\pm$	0.002 	&	1.0 	$\pm$	0.1 	&	0.10 	$\pm$	0.03 	&	12.7 	&5F0+F3	\\
17	&	52.182 	$\pm$	0.002 	&	0.8 	$\pm$	0.1 	&	0.40 	$\pm$	0.03 	&	10.5 	&2F0+2F3	\\
18	&	63.941 	$\pm$	0.002 	&	0.8 	$\pm$	0.1 	&	0.73 	$\pm$	0.03 	&	10.4 	&7F0	\\
19	&	19.576 	$\pm$	0.002 	&	0.7 	$\pm$	0.1 	&	0.30 	$\pm$	0.04 	&	8.1 	&4F0-F3	\\
20	&	1.782 	$\pm$	0.003 	&	1.0 	$\pm$	0.1 	&	0.74 	$\pm$	0.04 	&	7.7 	&g?	\\
21	&	28.709 	$\pm$	0.003 	&	0.7 	$\pm$	0.1 	&	0.63 	$\pm$	0.04 	&	7.6 	&5F0-F3	\\
22	&	1.715 	$\pm$	0.003 	&	1.0 	$\pm$	0.1 	&	0.60 	$\pm$	0.04 	&	7.1 	&g?	\\
23	&	71.767 	$\pm$	0.003 	&	0.6 	$\pm$	0.1 	&	0.38 	$\pm$	0.05 	&	7.0 	&6F0+F3	\\
24	&	1.646 	$\pm$	0.003 	&	0.9 	$\pm$	0.1 	&	0.56 	$\pm$	0.05 	&	6.7 	&g?	\\
25	&	1.850 	$\pm$	0.003 	&	0.8 	$\pm$	0.1 	&	0.83 	$\pm$	0.05 	&	6.0 	&g?	\\
26	&	73.077 	$\pm$	0.003 	&	0.5 	$\pm$	0.1 	&	0.98 	$\pm$	0.06 	&	5.7 	&8F0	\\
27	&	61.316 	$\pm$	0.004 	&	0.4 	$\pm$	0.1 	&	0.63 	$\pm$	0.06 	&	5.4 	&3F0+2F3	\\
28	&	80.892 	$\pm$	0.004 	&	0.4 	$\pm$	0.1 	&	0.80 	$\pm$	0.06 	&	5.2 	&7F0+F3	\\
\enddata
    \tablecomments{Among these frequencies, two peaks are independent frequencies, others are harmonic or combinations (denoted by $f_{i}$).}
\end{deluxetable}

\paragraph{TIC 255603395}

\begin{figure}[htp!]
\begin{center}
  \includegraphics[width=0.48\textwidth]{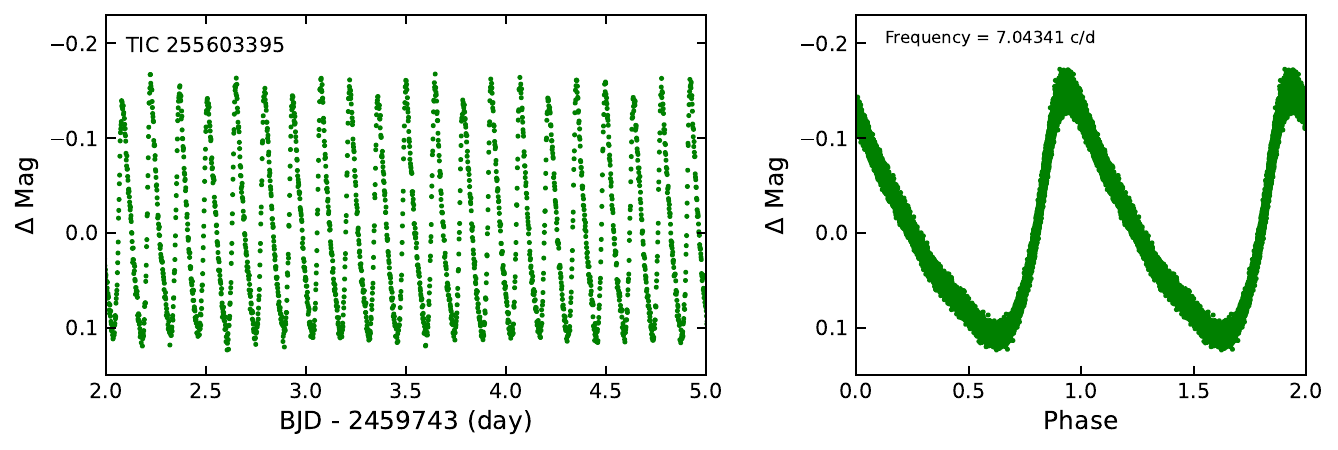}
  \includegraphics[width=0.482\textwidth]{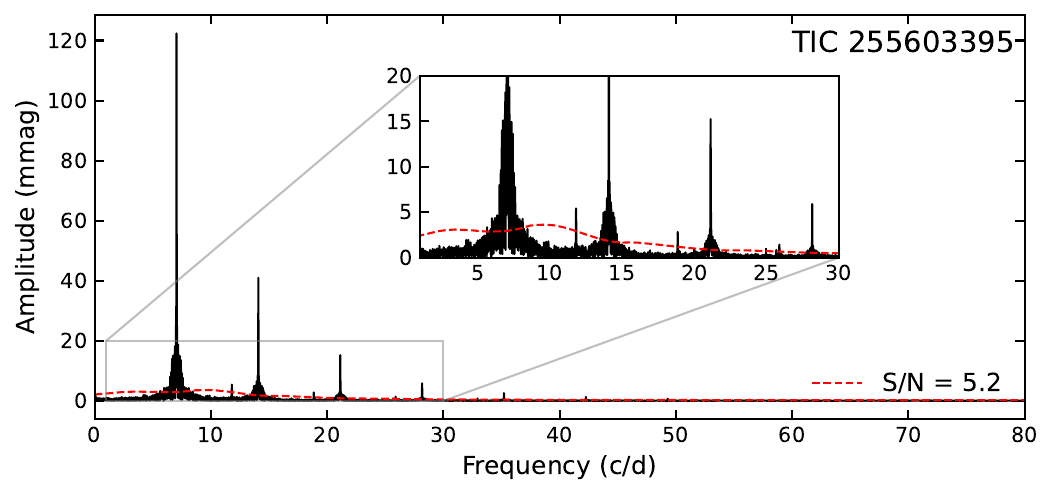}
  \caption{The upper left panel shows the light curve of TIC 255603395 covering 3 days. The upper right panel shows the phase diagram and the lower panel shows Fourier amplitude spectra.}
    \label{fig:TIC 255603395_light_fre}
\end{center}
\end{figure}

(ATO J283.3528+20.2049; $\alpha_{2000}$=18$^{h}$:53$^{m}$:24.672$^{s}$, $\delta_{2000}$=+20$\degr$:12$\arcmin $:17.835$\arcsec$) is a known \dsct star identified by \citet{Heinze2018}. We propose that this star could be a new double-mode HADS star. The TESS Space Telescope observed TIC 255603395 during Sector 53\&54 for a duration of 52.1 days, spanning from BJD 2459743.994812 to 2459796.132312, with a cadence of 2 minutes. After processing, we obtained a rectified light curve comprising 29426 data points. In Figure \ref{fig:TIC 255603395_light_fre}, the upper left panel displays a segment of the rectified light curve covering 3 days. The peak-to-peak amplitude of this star is $\sim$0.30 mag. Similarly to TIC 112682462, TIC 255603395 exhibits similarities to an RR Lyrae star.

A total of 31 frequencies were obtained by Fourier analysis. The two dominant modes in TIC 255603395 exhibit a frequency ratio of 0.596479, indicating that they correspond to the fundamental and second overtone radial modes \citep{Stellingwerf1979}. Similar to TIC 30977864, combination frequencies were detected using the same method. Hence, we present the fundamental frequency $f_{1}$ labeled as 'F0', the second overtone $f_{5}$ labeled as 'F2', the combination frequencies (such as $f_{6}$, $f_{8}$, $f_{9}$...$f_{30}$), and the harmonic frequencies (such as $f_{2}$, $f_{3}$, $f_{4}$...$f_{21}$) of the two radial modes in Table \ref{Tab112682462}. The ratio of $f_{0}$ / $f_{2}$ is determined to be 0.596479, slightly higher than the theoretical ratio. We infer that TIC 255603395 is a double-mode HADS star. It is worth noting that the first overtone frequencies were also not detected for some unknown reason. Additionally, we detected several non-radial frequencies (such as $f_{7}$, $f_{10}$, $f_{13}$...$f_{31}$) for which frequency combination was not possible. This target extracts low-order non-radial p mode pulsations as well as higher-order p modes, as measured by the values of these frequencies. The upper right panel of Figure \ref{fig:TIC 255603395_light_fre} shows the phase diagram of TIC 255603395. The lower panel of Figure \ref{fig:TIC 255603395_light_fre} shows the amplitude spectra.

\begin{deluxetable}{ccccccc}
\renewcommand\arraystretch{1.2}
\tabletypesize{\scriptsize}
\setlength\tabcolsep{1.5pt}
\tablewidth{0.5\textwidth}
\tablenum{4}
\tablecaption{The Pulsation Mode Frequencies in SC data of TIC 255603395. \label{Tab255603395}}
\tablehead{
\colhead{$f_{i}$}               &
\colhead{Frequency (day$^{-1}$)}  &
\colhead{Amplitude (mmag)}      &
\colhead{Phase (rad)}           &
\colhead{S/N}                   &
\colhead{Comment}               &
}
\startdata
1	&	7.04341 	$\pm$	0.00001 &	122.80 	$\pm$	0.06 	&	0.2637 	$\pm$	0.0002 	&	2084.3 	&F0	\\
2	&	14.08682 	$\pm$	0.00001 &	41.30 	$\pm$	0.05 	&	0.7024 	$\pm$	0.0004 	&	832.4 	&2F0	\\
3	&	21.13026 	$\pm$	0.00003 &	15.21 	$\pm$	0.05 	&	0.144 	$\pm$	0.001 	&	329.4 	&3F0	\\
4	&	28.17365 	$\pm$	0.00009 &	5.89 	$\pm$	0.05 	&	0.516 	$\pm$	0.002 	&	123.3 	&4F0	\\
5	&	11.8083	    $\pm$	0.0001 	&	5.61 	$\pm$	0.05 	&	0.560 	$\pm$	0.003 	&	106.8 	&F2	\\
6	&	18.8518  	$\pm$	0.0001 	&	2.84 	$\pm$	0.05 	&	0.887 	$\pm$	0.005 	&	62.0 	&F0+F2	\\
7	&	35.2170 	$\pm$	0.0001 	&	2.73 	$\pm$	0.05 	&	0.936 	$\pm$	0.005 	&	58.1 	&-	\\
8	&	4.7647	    $\pm$	0.0003 	&	1.78 	$\pm$	0.07 	&	0.18 	$\pm$	0.01 	&	27.1 	&F2-F0	\\
9	&	25.8955 	$\pm$	0.0003 	&	1.45 	$\pm$	0.05 	&	0.30 	$\pm$	0.01 	&	31.0 	&2F0+F2	\\
10	&	42.2600 	$\pm$	0.0003 	&	1.32 	$\pm$	0.04 	&	0.35 	$\pm$	0.01 	&	29.5 	&-	\\
11	&	2.2795  	$\pm$	0.0008 	&	0.90 	$\pm$	0.07 	&	0.28 	$\pm$	0.02 	&	13.1 	&2F0-F2	\\
12	&	32.9383 	$\pm$	0.0006 	&	0.75 	$\pm$	0.05 	&	0.75 	$\pm$	0.02 	&	15.9 	&3F0+F2	\\
13	&	49.3042 	$\pm$	0.0006 	&	0.70 	$\pm$	0.04 	&	0.75 	$\pm$	0.02 	&	15.6 	&-	\\
14	&	9.322 	    $\pm$	0.001 	&	0.55 	$\pm$	0.06 	&	0.73 	$\pm$	0.03 	&	9.7 	&3F0-F2	\\
15	&	12.384      $\pm$	0.001 	&	0.54 	$\pm$	0.05 	&	0.85 	$\pm$	0.03 	&	10.4 	&-	\\
16	&	25.683      $\pm$	0.001 	&	0.53 	$\pm$	0.05 	&	0.95 	$\pm$	0.02 	&	11.4 	&-	\\
17	&	11.410      $\pm$	0.001 	&	0.52 	$\pm$	0.05 	&	0.85 	$\pm$	0.03 	&	9.7 	&-	\\
18	&	12.450      $\pm$	0.001 	&	0.48 	$\pm$	0.05 	&	0.96 	$\pm$	0.03 	&	9.2 	&-	\\
19	&	13.186      $\pm$	0.001 	&	0.46 	$\pm$	0.05 	&	0.47 	$\pm$	0.03 	&	9.3 	&-	\\
20	&	39.981      $\pm$	0.001 	&	0.45 	$\pm$	0.05 	&	0.12 	$\pm$	0.03 	&	9.8 	&4F0+F2	\\
21	&	56.347      $\pm$	0.001 	&	0.41 	$\pm$	0.04 	&	0.18 	$\pm$	0.03 	&	9.4 	&8F0	\\
22	&	18.639      $\pm$	0.001 	&	0.39 	$\pm$	0.05 	&	0.08 	$\pm$	0.03 	&	8.4 	&-\\
23	&	25.227      $\pm$	0.001 	&	0.37 	$\pm$	0.05 	&	0.20 	$\pm$	0.04 	&	7.9 	&-	\\
24	&	24.429      $\pm$	0.001 	&	0.36 	$\pm$	0.05 	&	0.05 	$\pm$	0.04 	&	7.8 	&-	\\
25	&	32.270      $\pm$	0.001 	&	0.30 	$\pm$	0.05 	&	0.58 	$\pm$	0.04 	&	6.4 	&-	\\
26	&	20.231      $\pm$	0.001 	&	0.28 	$\pm$	0.05 	&	0.90 	$\pm$	0.05 	&	6.1 	&-	\\
27	&	31.472      $\pm$	0.001 	&	0.27 	$\pm$	0.05 	&	0.55 	$\pm$	0.05 	&	5.7 	&-	\\
28	&	16.365      $\pm$	0.001 	&	0.26 	$\pm$	0.05 	&	0.13 	$\pm$	0.05 	&	5.5 	&4F0-F2	\\
29	&	24.247      $\pm$	0.001 	&	0.26 	$\pm$	0.05 	&	0.88 	$\pm$	0.05 	&	5.6 	&-	\\
30	&	47.026      $\pm$	0.001 	&	0.26 	$\pm$	0.04 	&	0.53 	$\pm$	0.05 	&	5.9 	&5F0+F2	\\
31	&	31.099      $\pm$	0.002 	&	0.25 	$\pm$	0.05 	&	0.38 	$\pm$	0.06 	&	5.2 	&-	\\
\enddata
    \tablecomments{Among these frequencies, two peaks are independent frequencies, others are harmonic or combinations (denoted by $f_{i}$). '-' represents non-radial frequencies.}
\end{deluxetable}

\paragraph{TIC 281695001}

\begin{figure}[htp!]
\begin{center}
  \includegraphics[width=0.48\textwidth]{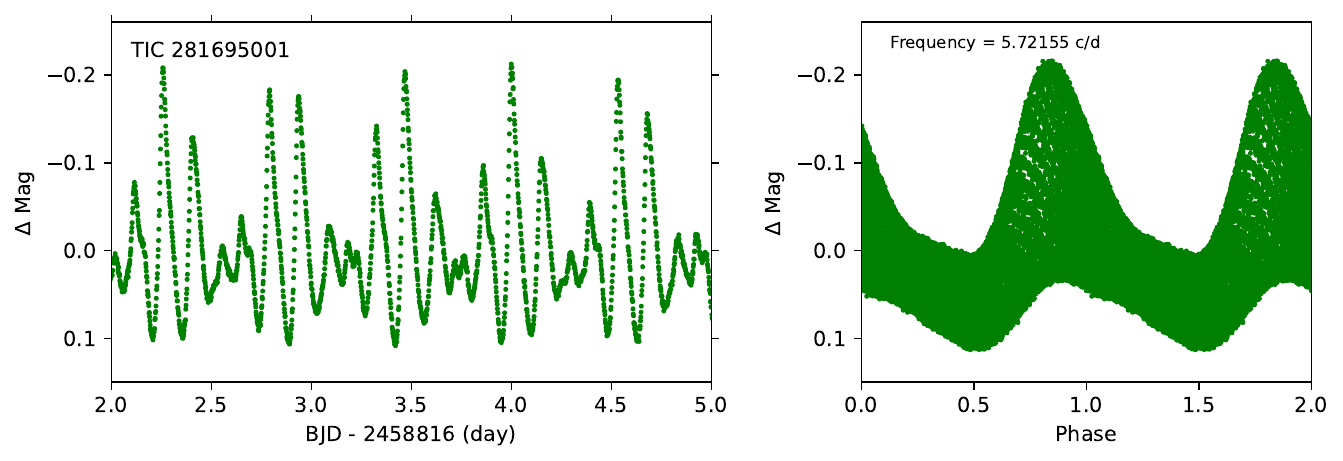}
  \includegraphics[width=0.482\textwidth]{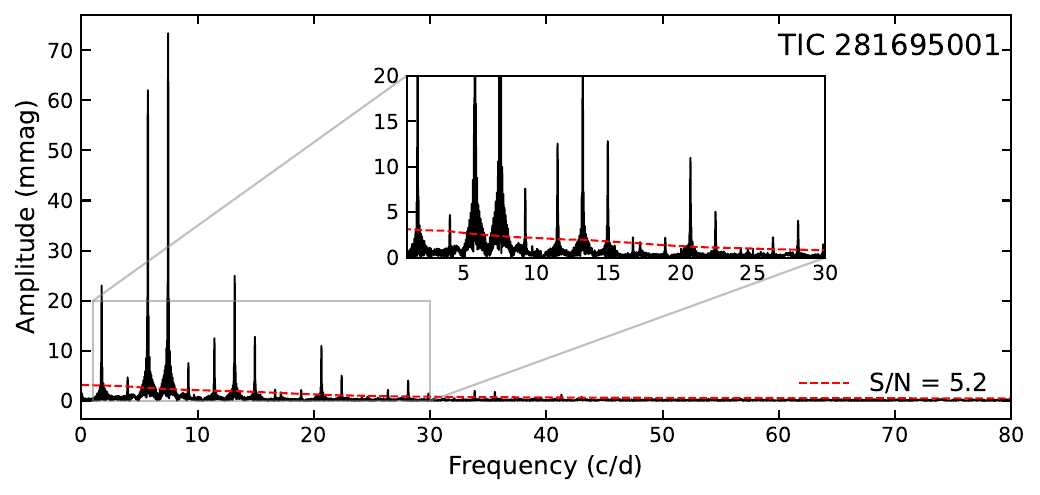}
  \caption{The upper left panel shows a portion of the light curve of TIC 281695001. The upper right panel shows the phase diagram and the lower panel shows Fourier amplitude spectra.}
    \label{fig:TIC 281695001_light_fre}
\end{center}
\end{figure}

(TYC 4342-685-1; $\alpha_{2000}$=04$^{h}$:53$^{m}$:46.556$^{s}$, $\delta_{2000}$=+68$\degr$:28$\arcmin $:26.397$\arcsec$)  is a known \dsct star confirmed by \citet{Cantat2018}. We first identified this star as a potential triple-mode HADS star. TIC 281695001 was observed for a duration of 24.9 days, from BJD 2458816.081346 to 2458840.992522 by TESS Sector 19. A rectified light curve consisting of 16687 data points was obtained. In the upper left panel of Figure \ref{fig:TIC 281695001_light_fre}, we present a portion of the light curve spanning 3 days. The figure shows a peak-to-peak amplitude of approximately 0.31 mag.

By Fourier transforming the spectrum of TIC 281695001, we detected a total of 49 frequencies. The three dominant modes in TIC 281695001 exhibit frequency ratios of 0.766606 and 0.621529, indicating that they correspond to the fundamental, first overtone, and second overtone radial modes, respectively. We consider the frequency of 5.72155(2)~day$^{-1}$ to be the fundamental frequency for this target since this frequency has a sufficiently large amplitude as well as a ratio to the first overtone. The upper right panel of Figure \ref{fig:TIC 281695001_light_fre} shows the phase diagram of TIC 281695001, folded by the fundamental frequency. The lower panel of Figure \ref{fig:TIC 281695001_light_fre} presents the amplitude spectra. Table \ref{Tab281695001} lists the fundamental frequency $f_{2}$ labeled as 'F0', the first overtone $f_{1}$ labeled as 'F1', the second overtone $f_{8}$ labeled as 'F2', as well as the combination frequencies (such as $f_{3}$, $f_{4}$, $f_{7}$...$f_{49}$) and the harmonic frequencies (such as $f_{5}$, $f_{6}$, $f_{9}$...$f_{41}$) of the three radial modes. Additionally, we detected two non-radial frequencies ($f_{25}$, $f_{47}$), $f_{25}$ might be a low-order non-radial p mode and $f_{47}$ a high-order non-radial p mode. Notably, the amplitude of the first overtone in this target is larger than that of the fundamental frequency. The driving and damping mechanisms for \dsct stars, specifically the mode selection processes, are still not fully understood. Consequently, further investigations are necessary to elucidate the underlying reasons for this phenomenon.

\begin{deluxetable}{ccccccc}
\renewcommand\arraystretch{1.2}
\tabletypesize{\scriptsize}
\setlength\tabcolsep{1.3pt}
\tablewidth{0.5\textwidth}
\tablenum{5}
\tablecaption{The Pulsation Mode Frequencies in SC data of TIC 281695001. \label{Tab281695001}}
\tablehead{
\colhead{$f_{i}$}               &
\colhead{Frequency (day$^{-1}$)}  &
\colhead{Amplitude (mmag)}      &
\colhead{Phase (rad)}           &
\colhead{S/N}                   &
\colhead{Comment}               &
}
\startdata
1	&	7.46348 $\pm$	0.00001 &	73.27 	$\pm$	0.04 	&	0.6433 	$\pm$	0.0002 	&	1774.6 	&F1	\\
2	&	5.72155 $\pm$	0.00002 &	62.14 	$\pm$	0.04 	&	0.5116 	$\pm$	0.0002 	&	1460.1 	&F0	\\
3	&	13.18498$\pm$	0.00003 &	24.89 	$\pm$	0.04 	&	0.3239 	$\pm$	0.0005 	&	666.4 	&F0+F1	\\
4	&	1.74220 $\pm$	0.00005 &	22.73 	$\pm$	0.05 	&	0.1485 	$\pm$	0.0007 	&	488.1 	&F1-F0	\\
5	&	14.92708$\pm$	0.00006 &	12.70 	$\pm$	0.04 	&	0.5458 	$\pm$	0.0009 	&	350.3 	&2F1	\\
6	&	11.44309$\pm$	0.00007 &	12.33 	$\pm$	0.04 	&	0.183 	$\pm$	0.001 	&	318.3 	&2F0	\\
7	&	20.64848$\pm$	0.00007 &	11.00 	$\pm$	0.03 	&	0.044 	$\pm$	0.001 	&	326.8 	&F0+2F1	\\
8	&	9.2056 	$\pm$	0.0001 	&	7.88 	$\pm$	0.04 	&	0.857 	$\pm$	0.001 	&	195.5 	&F2	\\
9	&	22.3909 $\pm$	0.0001 	&	5.08 	$\pm$	0.03 	&	0.321 	$\pm$	0.002 	&	151.7 	&3F1	\\
10	&	3.9799 	$\pm$	0.0002 	&	4.87 	$\pm$	0.04 	&	0.463 	$\pm$	0.002 	&	109.8 	&2F0-F1	\\
11	&	28.1119 $\pm$	0.0001 	&	4.04 	$\pm$	0.03 	&	0.841 	$\pm$	0.002 	&	131.7 	&F0+3F1	\\
12	&	26.3693 $\pm$	0.0003 	&	2.26 	$\pm$	0.03 	&	0.686 	$\pm$	0.004 	&	72.0 	&2F0+2F1	\\
13	&	18.9060 $\pm$	0.0003 	&	2.22 	$\pm$	0.03 	&	0.021 	$\pm$	0.004 	&	65.3 	&2F0+F1	\\
14	&	33.8332 $\pm$	0.0003 	&	2.10 	$\pm$	0.03 	&	0.466 	$\pm$	0.004 	&	70.2 	&2F0+3F1	\\
15	&	16.6695 $\pm$	0.0003 	&	2.08 	$\pm$	0.04 	&	0.634 	$\pm$	0.005 	&	58.4 	&F1+F2	\\
16	&	17.1646 $\pm$	0.0004 	&	1.91 	$\pm$	0.04 	&	0.847 	$\pm$	0.005 	&	54.4 	&3F0	\\
17	&	35.5753 $\pm$	0.0004 	&	1.65 	$\pm$	0.03 	&	0.677 	$\pm$	0.005 	&	54.7 	&F0+4F1	\\
18	&	29.8542 $\pm$	0.0004 	&	1.39 	$\pm$	0.03 	&	0.147 	$\pm$	0.007 	&	45.4 	&F0+2F1+F2	\\
19	&	41.2970 $\pm$	0.0005 	&	1.20 	$\pm$	0.03 	&	0.258 	$\pm$	0.007 	&	40.9 	&3F0+2F1+F2	\\
20	&	24.6270 $\pm$	0.0006 	&	1.11 	$\pm$	0.03 	&	0.695 	$\pm$	0.009 	&	33.5 	&3F0+F1	\\
21	&	9.7014	$\pm$	0.0008 	&	1.09 	$\pm$	0.04 	&	0.12 	$\pm$	0.01 	&	27.3 	&F0+3F1-2F2	\\
22	&	24.1334 $\pm$	0.0007 	&	0.97 	$\pm$	0.03 	&	0.45 	$\pm$	0.01 	&	29.6 	&2F1+F2	\\
23	&	39.5544 $\pm$	0.0009 	&	0.69 	$\pm$	0.03 	&	0.19 	$\pm$	0.01 	&	23.6 	&4F0+F1+F2	\\
24	&	10.94696$\pm$	0.001 	&	0.63 	$\pm$	0.04 	&	0.30 	$\pm$	0.01 	&	16.2 	&2F2-F1	\\
25	&	6.302 	$\pm$	0.001 	&	0.63 	$\pm$	0.04 	&	0.10 	$\pm$	0.02 	&	14.9 	&-	\\
26	&	43.038 	$\pm$	0.001 	&	0.61 	$\pm$	0.03 	&	0.48 	$\pm$	0.01 	&	20.8 	&2F0+3F1+F2	\\
27	&	48.759 	$\pm$	0.001 	&	0.58 	$\pm$	0.03 	&	0.07 	$\pm$	0.01 	&	20.3 	&4F0+F1+2F2	\\
28	&	47.017 	$\pm$	0.001 	&	0.55 	$\pm$	0.03 	&	0.96 	$\pm$	0.01 	&	19.3 	&3F0+4F1	\\
29	&	2.235 	$\pm$	0.001 	&	0.55 	$\pm$	0.05 	&	0.72 	$\pm$	0.02 	&	12.0 	&3F0-2F1	\\
30	&	32.089 	$\pm$	0.001 	&	0.45 	$\pm$	0.03 	&	0.42 	$\pm$	0.02 	&	15.0 	&3F0+2F1	\\
31	&	37.315 	$\pm$	0.001 	&	0.44 	$\pm$	0.03 	&	0.99 	$\pm$	0.02 	&	14.6 	&F0+3F1+F2	\\
32	&	31.597 	$\pm$	0.001 	&	0.42 	$\pm$	0.03 	&	0.25 	$\pm$	0.02 	&	14.0 	&3F1+F2	\\
33	&	45.275 	$\pm$	0.001 	&	0.35 	$\pm$	0.03 	&	0.87 	$\pm$	0.02 	&	12.4 	&4F0+3F1	\\
34	&	18.411 	$\pm$	0.002 	&	0.35 	$\pm$	0.03 	&	0.00 	$\pm$	0.03 	&	10.2 	&F0-2F1+3F2	\\
35	&	15.422 	$\pm$	0.002 	&	0.34 	$\pm$	0.04 	&	0.87 	$\pm$	0.03 	&	9.5 	&4F0-F1	\\
36	&	5.226 	$\pm$	0.002 	&	0.33 	$\pm$	0.04 	&	0.42 	$\pm$	0.04 	&	7.7 	&F1+F2-2F0	\\
37	&	54.481 	$\pm$	0.002 	&	0.33 	$\pm$	0.03 	&	0.76 	$\pm$	0.02 	&	11.1 	&4F0+3F1+F2	\\
38	&	7.957 	$\pm$	0.002 	&	0.32 	$\pm$	0.04 	&	0.27 	$\pm$	0.04 	&	7.8 	&4F0-2F1	\\
39	&	37.812 	$\pm$	0.002 	&	0.30 	$\pm$	0.03 	&	0.16 	$\pm$	0.03 	&	10.1 	&4F0-2F1+4F2	\\
40	&	52.740 	$\pm$	0.002 	&	0.29 	$\pm$	0.03 	&	0.65 	$\pm$	0.03 	&	9.8 	&4F0+4F1	\\
41	&	22.888 	$\pm$	0.002 	&	0.28 	$\pm$	0.03 	&	0.40 	$\pm$	0.03 	&	8.5 	&4F0	\\
42	&	50.504 	$\pm$	0.002 	&	0.27 	$\pm$	0.03 	&	0.25 	$\pm$	0.03 	&	9.1 	&4F0+3F2	\\
43	&	61.947 	$\pm$	0.002 	&	0.25 	$\pm$	0.03 	&	0.55 	$\pm$	0.03 	&	8.6 	&4F0+4F1+F2	\\
44	&	56.224 	$\pm$	0.002 	&	0.24 	$\pm$	0.03 	&	0.83 	$\pm$	0.03 	&	8.0 	&4F0+4F1+2F2	\\
45	&	39.057 	$\pm$	0.003 	&	0.21 	$\pm$	0.03 	&	0.08 	$\pm$	0.04 	&	7.2 	&3F0-2F1+4F2	\\
46	&	13.675 	$\pm$	0.004 	&	0.19 	$\pm$	0.04 	&	0.96 	$\pm$	0.06 	&	5.2 	&2F0+4F1-3F2	\\
47	&	60.204 	$\pm$	0.003 	&	0.19 	$\pm$	0.03 	&	0.44 	$\pm$	0.04 	&	6.5 	&-	\\
48	&	44.775 	$\pm$	0.003 	&	0.16 	$\pm$	0.03 	&	0.86 	$\pm$	0.05 	&	5.7 	&4F0-2F1+4F2	\\
49	&	30.348 	$\pm$	0.004 	&	0.16 	$\pm$	0.03 	&	0.43 	$\pm$	0.05 	&	5.4 	&4F0+F1	\\
   \enddata
    \tablecomments{Among these frequencies, two peaks are independent frequencies, others are harmonic or combinations (denoted by $f_{i}$). '-' represents non-radial frequencies.}
\end{deluxetable}

\paragraph{TIC 353050847}

\begin{figure}[htp!]
\begin{center}
  \includegraphics[width=0.48\textwidth]{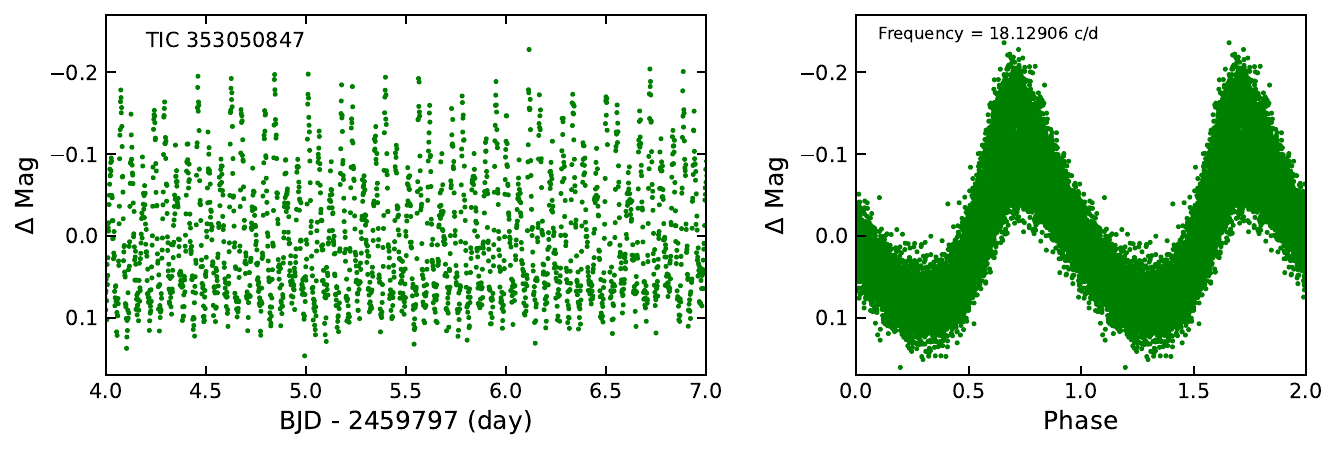}
  \includegraphics[width=0.482\textwidth]{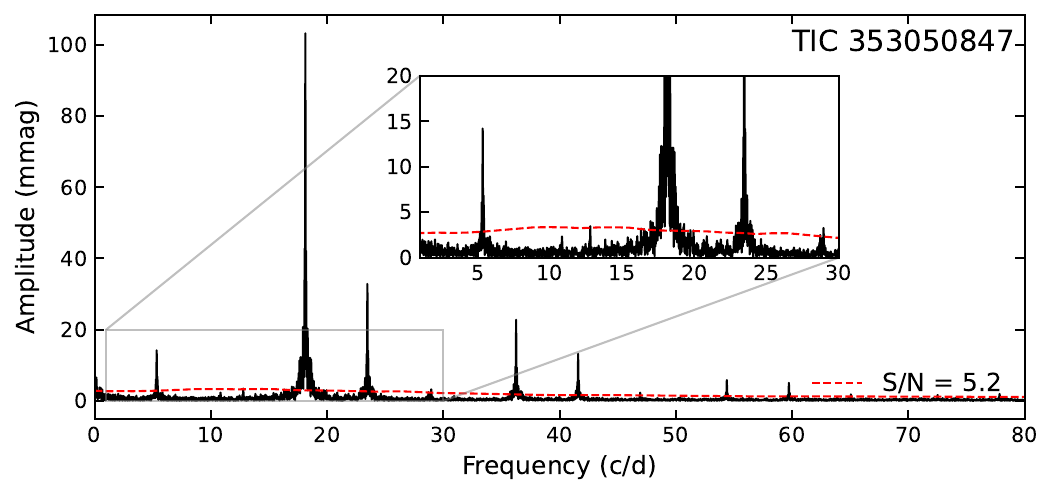}
  \caption{The upper left panel shows a portion of the light curve of TIC 353050847. The upper right panel shows the phase diagram and the lower panel shows Fourier amplitude spectra.}
    \label{fig:TIC 353050847_light_fre}
\end{center}
\end{figure}

(UCAC4 481-131734; $\alpha_{2000}$=21$^{h}$:37$^{m}$:34.109$^{s}$, $\delta_{2000}$=+06$\degr$:05$\arcmin $:05.732$\arcsec$) is a newly identified triple-mode HADS star based on our analysis. TIC 353050847 was observed from BJD 2459797.100352 to 2459824.268491 during Sector 55 for 27.1 days. After processing, a rectified light curve consisting of 13395 points was obtained. Figure \ref{fig:TIC 353050847_light_fre} upper left panel shows a portion of the light curve covering 3 days. The figure reveals a peak-to-peak amplitude of $\sim$0.31 mag. The low-resolution spectral parameters obtained from LAMOST are T$_{\rm eff}$ = 7413.80 $\pm$ 168.67 (K), log $g$ = 4.358 $\pm$ 0.278, and [Fe/H] = -0.860 $\pm$ 0.182.

We detected a total of 14 frequencies by Fourier analysis of TIC 353050847. The three dominant modes in TIC 353050847 exhibit frequency ratios of 0.772596 and 0.626068, which correspond to the fundamental, first overtone, and second overtone radial modes, respectively. Table \ref{Tab353050847} lists the fundamental frequency $f_{1}$ labeled as 'F0', the first overtone $f_{2}$ labeled as 'F1', the second overtone $f_{9}$ labeled as 'F2', as well as the combination frequencies (such as $f_{4}$, $f_{5}$, $f_{7}$...$f_{13}$) and the harmonic frequencies (such as $f_{3}$, $f_{6}$, $f_{12}$, $f_{14}$) of the three radial modes. The upper right panel of Figure \ref{fig:TIC 353050847_light_fre} shows the phase diagram of TIC 353050847. The lower panel of Figure \ref{fig:TIC 353050847_light_fre} shows the amplitude spectra. Additionally, we detected two frequencies ($f_{10}$, $f_{11}$) that might be the low-order non-radial p modes. The amplitude of F2 is relatively small, making it difficult to accurately determine the frequency as the second overtone, suggesting that this target may indeed be a triple-mode HADS star.

\begin{deluxetable}{ccccccc}
\renewcommand\arraystretch{1.2}
\tabletypesize{\scriptsize}
\setlength\tabcolsep{2.0pt}
\tablewidth{0.5\textwidth}
\tablenum{6}
\tablecaption{The Pulsation Mode Frequencies in SC data of TIC 353050847. \label{Tab353050847}}
\tablehead{
\colhead{$f_{i}$}               &
\colhead{Frequency (day$^{-1}$)}  &
\colhead{Amplitude (mmag)}      &
\colhead{Phase (rad)}           &
\colhead{S/N}                   &
\colhead{Comment}               &
}
\startdata
1	&	18.12906 	$\pm$	0.00006 	&	103.3 	$\pm$	0.3 	&	0.9345 	$\pm$	0.0009 	&	346.3 	&F0	\\
2	&	23.4651 	$\pm$	0.0001 	    &	33.1 	$\pm$	0.3 	&	0.710 	$\pm$	0.002 	&	118.5 	&F1	\\
3	&	36.2583 	$\pm$	0.0002 	    &	22.9 	$\pm$	0.3 	&	0.986 	$\pm$	0.003 	&	84.5 	&2F0	\\
4	&	5.3363 	    $\pm$	0.0005 	    &	14.8 	$\pm$	0.4 	&	0.740 	$\pm$	0.008 	&	36.3 	&F1-F0	\\
5	&	41.5939 	$\pm$	0.0004 	    &	13.1 	$\pm$	0.3 	&	0.761 	$\pm$	0.006 	&	47.7 	&F0+F1	\\
6	&	54.38587 	$\pm$	0.0009 	    &	5.8 	$\pm$	0.3 	&	0.06 	$\pm$	0.01 	&	21.4 	&3F0	\\
7	&	59.723   	$\pm$	0.001 	    &	5.0 	$\pm$	0.3 	&	0.80 	$\pm$	0.01 	&	18.5 	&2F0+F1	\\
8	&	12.791   	$\pm$	0.002 	    &	3.4 	$\pm$	0.3 	&	0.29 	$\pm$	0.03 	&	10.1 	&2F0-F1	\\
9	&	28.957   	$\pm$	0.001 	    &	3.1 	$\pm$	0.3 	&	0.91 	$\pm$	0.02 	&	11.4 	&F2	\\
10	&	28.730   	$\pm$	0.002 	    &	2.6 	$\pm$	0.3 	&	0.45 	$\pm$	0.03 	&	9.8 	&-	\\
11	&	22.272   	$\pm$	0.002 	    &	2.3 	$\pm$	0.3 	&	0.27 	$\pm$	0.03 	&	8.0 	&-	\\
12	&	46.929   	$\pm$	0.002 	    &	2.3 	$\pm$	0.3 	&	0.51 	$\pm$	0.03 	&	8.4 	&2F1	\\
13	&	65.057   	$\pm$	0.002 	    &	1.8 	$\pm$	0.3 	&	0.59 	$\pm$	0.04 	&	6.9 	&F0+2F1	\\
14	&	72.515   	$\pm$	0.003 	    &	1.7 	$\pm$	0.3 	&	0.16 	$\pm$	0.05 	&	6.1 	&4F0	\\
   \enddata
    \tablecomments{Among these frequencies, two peaks are independent frequencies, others are harmonic or combinations (denoted by $f_{i}$). '-' represents non-radial frequencies.}
\end{deluxetable}

\paragraph{TIC 387379145}

\begin{figure}[htp!]
\begin{center}
  \includegraphics[width=0.48\textwidth]{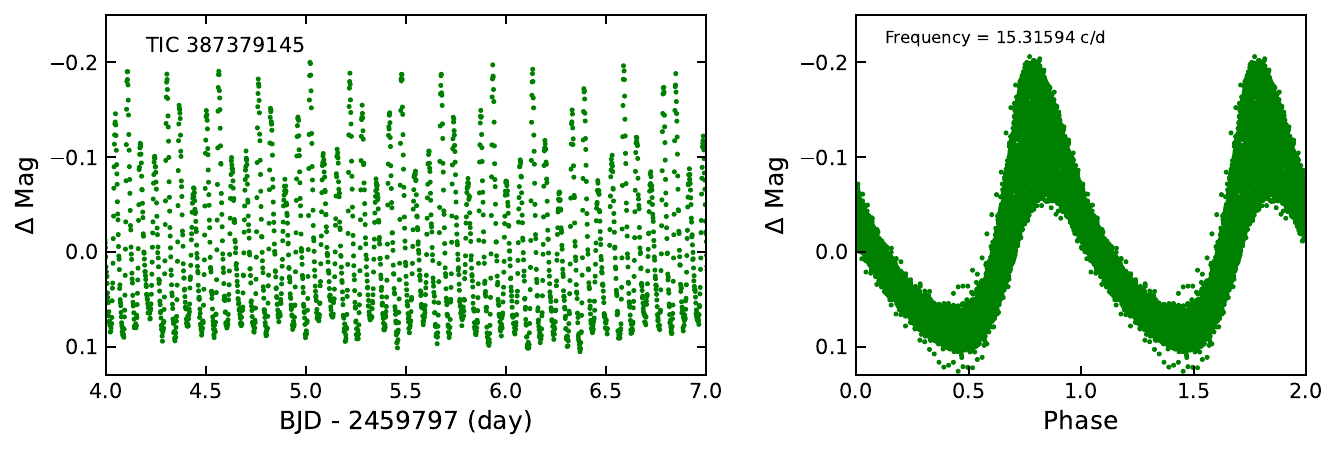}
  \includegraphics[width=0.482\textwidth]{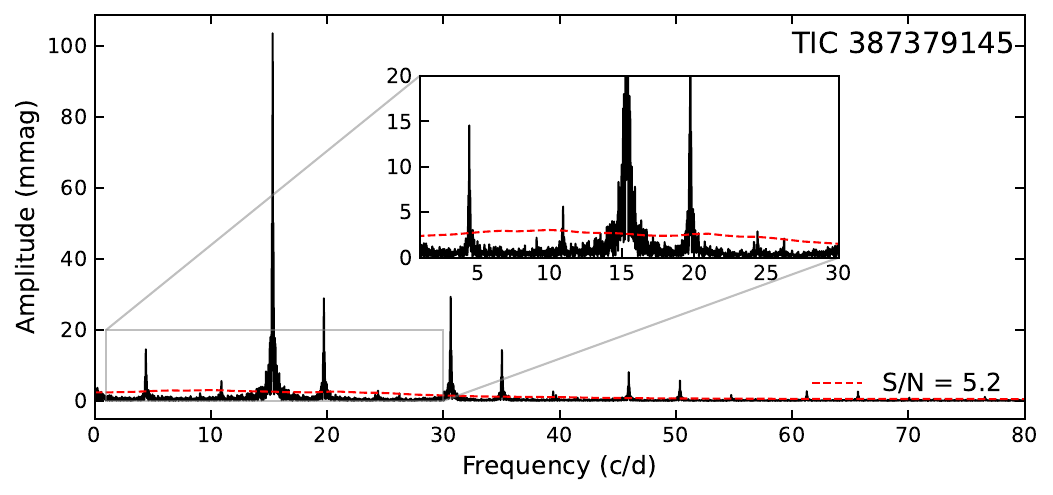}
  \caption{The upper left panel shows a portion of the light curve of TIC 387379145. The upper right panel shows the phase diagram and the lower panel shows Fourier amplitude spectra.}
    \label{fig:TIC 387379145_light_fre}
\end{center}
\end{figure}

(CRTS J211314.3+033022; $\alpha_{2000}$=21$^{h}$:13$^{m}$:14.276$^{s}$, $\delta_{2000}$=+03$\degr$:30$\arcmin $:21.918$\arcsec$) is a known \dsct star certified by \citet{Drake2014}. Through our analysis, we first identified this star might be a new triple-mode HADS star. TESS observed TIC 387379145 during Sector 55 for 27.1 days from Barycentric Julian Date (BJD) 2459797.100515 to 2459824.268654 including 13185 data points. The upper left panel of Figure \ref{fig:TIC 387379145_light_fre} shows a portion of the light curve of this star covering 3 days. From this figure, the peak-to-peak amplitude of this star obtained from the rectified light curve is about 0.30 mag.

A total of 40 frequencies were detected through Fourier analysis. The three high-amplitude independent modes of TIC 387379145 have a frequency ratio of 0.776588 and 0.628124, respectively, identifying them as the fundamental, first overtone, and second overtone radial modes. Therefore, the fundamental frequency $f_{1}$ with 'F0', first overtone $f_{3}$ with 'F1', second overtone $f_{14}$ with 'F2' and the combination frequencies (i.e., $f_{4}$, $f_{5}$, $f_{7}$...$f_{40}$) and harmonics frequencies (i.e., $f_{2}$, $f_{6}$, $f_{12}$, $f_{30}$) of the three radial modes are listed in Table \ref{Tab387379145}. The upper right panel of Figure \ref{fig:TIC 387379145_light_fre} shows the phase diagram of TIC 387379145, folded by the fundamental frequency F0 = 15.31594(2)~day$^{-1}$. The lower panel of Figure \ref{fig:TIC 387379145_light_fre} shows the amplitude spectra. There is no non-radial mode found for this star.

\begin{deluxetable}{ccccccc}
\renewcommand\arraystretch{1.2}
\tabletypesize{\scriptsize}
\setlength\tabcolsep{1.5pt}
\tablewidth{0.5\textwidth}
\tablenum{7}
\tablecaption{The Pulsation Mode Frequencies in SC data of TIC 387379145. \label{Tab387379145}}
\tablehead{
\colhead{$f_{i}$}               &
\colhead{Frequency (day$^{-1}$)}  &
\colhead{Amplitude (mmag)}      &
\colhead{Phase (rad)}           &
\colhead{S/N}                   &
\colhead{Comment}               &
}
\startdata
1	&	15.31594 $\pm$	0.00002 &	103.7 	$\pm$	0.1 	&	0.568 	$\pm$	0.0004 	&	893.0 	&F0		\\
2	&	30.63176 $\pm$	0.00005 &	29.7 	$\pm$	0.1 	&	0.266 	$\pm$	0.0009 	&	374.2 	&2F0		\\
3	&	19.72208 $\pm$	0.00007 &	29.0 	$\pm$	0.1 	&	0.988 	$\pm$	0.001 	&	286.3 	&F1		\\
4	&	4.4059 	$\pm$	0.0002 	&	15.0 	$\pm$	0.2 	&	0.364 	$\pm$	0.003 	&	81.8 	&F1-F0		\\
5	&	35.0381 $\pm$	0.0001 	&	14.7 	$\pm$	0.1 	&	0.674 	$\pm$	0.001 	&	193.2 	&F0+F1		\\
6	&	45.9476 $\pm$	0.0001 	&	8.5 	$\pm$	0.1 	&	0.971 	$\pm$	0.002 	&	121.6 	&3F0	\\
7	&	50.3539 $\pm$	0.0002 	&	6.2 	$\pm$	0.1 	&	0.372 	$\pm$	0.003 	&	88.8 	&2F0+F1		\\
8	&	10.9096 $\pm$	0.0004 	&	6.0 	$\pm$	0.1 	&	0.319 	$\pm$	0.007 	&	41.4 	&2F0-F1		\\
12	&	61.2641 $\pm$	0.0004 	&	2.9 	$\pm$	0.1 	&	0.705 	$\pm$	0.007 	&	44.5 	&4F0		\\
13	&	65.6693 $\pm$	0.0004 	&	2.8 	$\pm$	0.1 	&	0.099 	$\pm$	0.007 	&	44.7 	&3F0+F1		\\
14	&	24.3836 $\pm$	0.0006 	&	2.7 	$\pm$	0.1 	&	0.78 	$\pm$	0.01 	&	31.0 	&F2		\\
15	&	39.4437 $\pm$	0.0005 	&	2.5 	$\pm$	0.1 	&	0.11 	$\pm$	0.01 	&	34.6 	&2F1		\\
17	&	1.255 	$\pm$	0.001 	&	2.4 	$\pm$	0.2 	&	0.97 	$\pm$	0.02 	&	10.8 	&4F2-5F0-F1		\\
19	&	26.2240 $\pm$	0.0007 	&	2.3 	$\pm$	0.1 	&	0.02 	$\pm$	0.01 	&	27.8 	&3F0-F1		\\
20	&	54.7596 $\pm$	0.0007 	&	1.8 	$\pm$	0.1 	&	0.80 	$\pm$	0.01 	&	27.3 	&F0+2F1	\\
22	&	24.126 	$\pm$	0.001 	&	1.6 	$\pm$	0.1 	&	0.44 	$\pm$	0.01 	&	18.1 	&2F1-F0		\\
24	&	14.161 	$\pm$	0.001 	&	1.5 	$\pm$	0.1 	&	0.57 	$\pm$	0.02 	&	11.9 	&F0-5F1+4F2		\\
26	&	39.698 	$\pm$	0.001 	&	1.4 	$\pm$	0.1 	&	0.50 	$\pm$	0.01 	&	19.4 	&F0+F2		\\
29	&	9.063 	$\pm$	0.002 	&	1.2 	$\pm$	0.2 	&	0.10 	$\pm$	0.04 	&	7.8 	&F2-F0		\\
30	&	76.578 	$\pm$	0.001 	&	1.1 	$\pm$	0.1 	&	0.44 	$\pm$	0.01 	&	17.7 	&5F0		\\
31	&	70.076 	$\pm$	0.001 	&	1.0 	$\pm$	0.1 	&	0.51 	$\pm$	0.01 	&	16.3 	&2F0+2F1		\\
33	&	28.265 	$\pm$	0.001 	&	0.9 	$\pm$	0.1 	&	0.83 	$\pm$	0.02 	&	11.3 	&5F1-3F0-F2		\\
34	&	41.541 	$\pm$	0.001 	&	0.9 	$\pm$	0.1 	&	0.75 	$\pm$	0.02 	&	12.9 	&4F0-F1		\\
35	&	24.496 	$\pm$	0.001 	&	0.9 	$\pm$	0.1 	&	0.41 	$\pm$	0.03 	&	10.2 	&6F0+4F1-6F2		\\
36	&	27.989 	$\pm$	0.002 	&	0.6 	$\pm$	0.1 	&	0.41 	$\pm$	0.04 	&	7.3 	&6F2-6F1		\\
37	&	44.106 	$\pm$	0.002 	&	0.5 	$\pm$	0.1 	&	0.86 	$\pm$	0.04 	&	7.3 	&F1+F2		\\
38	&	55.016 	$\pm$	0.002 	&	0.5 	$\pm$	0.1 	&	0.17 	$\pm$	0.04 	&	7.2 	&2F0+F2		\\
39	&	59.418 	$\pm$	0.002 	&	0.5 	$\pm$	0.1 	&	0.63 	$\pm$	0.04 	&	7.3 	&F0+F1+F2		\\
40	&	28.227 	$\pm$	0.003 	&	0.5 	$\pm$	0.1 	&	0.50 	$\pm$	0.05 	&	5.6 	&5F1-3F0-F2		\\
   \enddata
    \tablecomments{Among these frequencies, two peaks are independent frequencies, others are harmonic or combinations (denoted by $f_{i}$).}
\end{deluxetable}

\section{The distribution of the period ratio}

\begin{figure}[h!]
\begin{center}
  \includegraphics[width=0.48\textwidth]{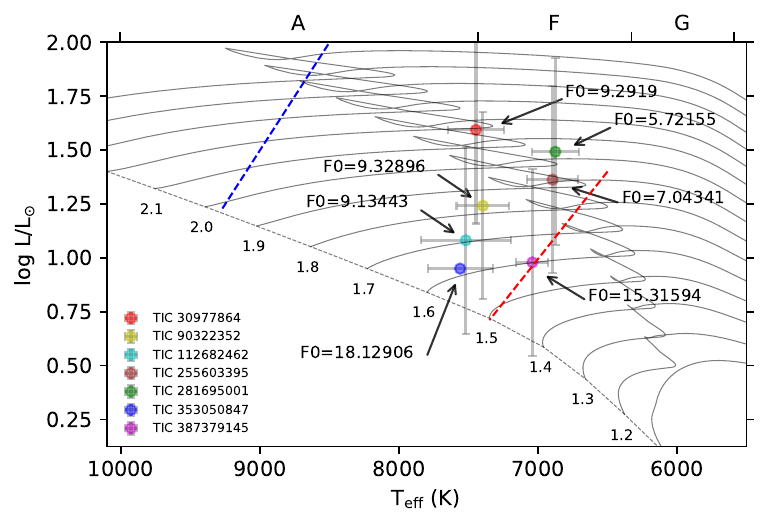}
  \caption{HR diagram of the newly discovered HADS. The different color dots represent the different HADS. The theoretical instability strip boundaries from \citet{Murphy2019}. The evolutionary tracks were calculated in MESA v10108 with X = 0.71 and Z = 0.01. Additional comprehensive information about these tracks can be found in \citet{Murphy2019}.}
    \label{fig:HR_HADS}
\end{center}
\end{figure}

\begin{figure}[h!]
\begin{center}
  \includegraphics[width=0.48\textwidth]{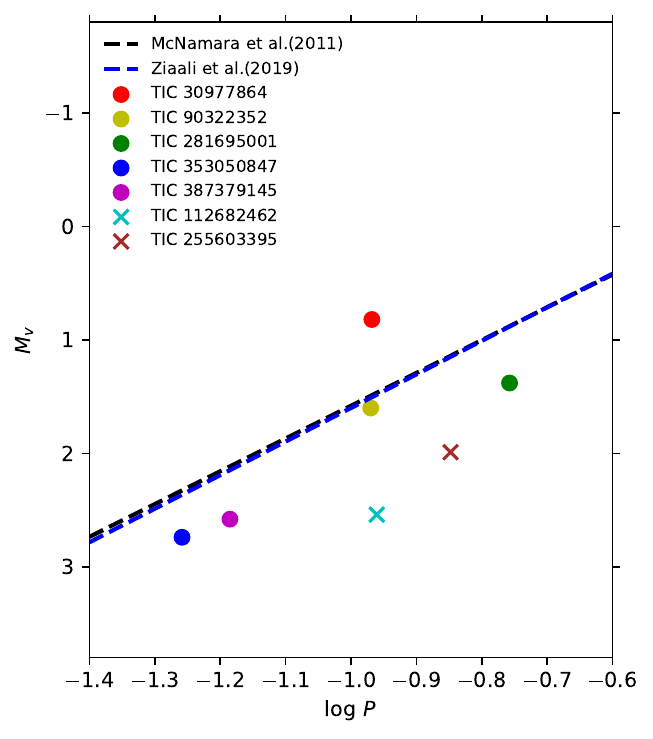}
  \caption{Period-luminosity relation of seven HADS stars. The black dashed line is the P-L relation for High-amplitude \dsct stars derived by \citet{2011AJ....142..110M}, while the P-L relation obtained by \citet{2019MNRAS.486.4348Z} is shown by the blue dashed line.}
    \label{fig:PLR}
\end{center}
\end{figure}

\setcounter{table}{7}
\begin{table*}[!t]\scriptsize
\renewcommand\arraystretch{1.2}
\setlength\tabcolsep{5.5pt}
\caption{TIC number, positions,$V$mag, effective temperature (T$_{\rm eff}$), Mass ($M/M_{\odot}$), Radius ($R/R_{\odot}$), surface gravity ($\log g$), Density(g/cm$^{3}$), Luminosity($L/L_{\odot}$), Distance(pc), and $M_v$ for 7 new identified HADS stars. \label{HADS_para}}
\tabletypesize{\scriptsize}
\centering
\begin{tabular}{ccccccccccccc}
\hline
\hline
TIC& RA (deg) & Dec (deg) &$V$mag&T$_{\rm eff}$ (K)&$M/M_{\odot}$&$R/R_{\odot}$&$\log g$ (dex)&Density(g/cm$^{3}$)&$L/L_{\odot}$ &Distance(pc)&$M_v$\\
\hline
30977864  	&	287.810212	&	-34.896129	   &	14.42 	       &   7447    $\pm$	203 	&	1.71                	&	3.76                 	&	3.52                	&  0.045                   &	39.22 	$\pm$	0.48 	&	5259    $\pm$	1120    &	0.82 	\\
90322352    &	276.401048	&	-42.226624	   &	11.52 	       &   7397    $\pm$	190 	&	1.69 	$\pm$	0.27 	&	2.54 	$\pm$	0.18 	&	3.86 	$\pm$	0.10 	&  0.145   $\pm$	0.041  &	17.43 	$\pm$	1.95 	&	963 	$\pm$	50 	    &	1.60 	\\
112682462  	&	269.602262	&	-41.841784     &	13.12 	       &   7520    $\pm$	322 	&	1.74 	$\pm$	0.31 	&	2.05 	$\pm$	0.20 	&	4.06 	$\pm$	0.13 	&  0.287   $\pm$	0.109  &	12.06 	$\pm$	1.65 	&	1306 	$\pm$	77 	    &	2.54 	\\
255603395  	&	283.352799	&	20.204954      &	12.59 	       &   6896    $\pm$	182 	&	1.50 	$\pm$	0.25 	&	3.36 	$\pm$	0.27 	&	3.56 	$\pm$	0.11 	&  0.056   $\pm$	0.017  &	23.05 	$\pm$	3.33 	&	1317 	$\pm$	84 	    &	1.99 	\\
281695001   &	73.443985	&	68.473999      &	11.66 	       &   6876    $\pm$	168 	&	1.49 	$\pm$	0.27 	&	3.93 	$\pm$	0.24 	&	3.42 	$\pm$	0.10 	&  0.035   $\pm$	0.009  &	31.03 	$\pm$	2.36 	&	1137 	$\pm$	33 	    &	1.38 	\\
353050847   &	324.392118	&	6.084925       &	14.79 	       &   7561    $\pm$	234 	&	1.75                	&	1.73                 	&	4.20 	            	&  0.468                   &	8.91                	&	2576 	$\pm$	263	    &	2.74 	\\
387379145   &	318.309483	&	3.506088       &	12.63 	       &   7041    $\pm$	115 	&	1.55 	$\pm$	0.27 	&	2.07 	$\pm$	0.13 	&	3.99 	$\pm$	0.09 	&  0.245   $\pm$	0.065  &	9.53 	$\pm$	1.02 	&	1024 	$\pm$	46 	    &	2.58 	\\
\hline
\end{tabular}
\end{table*}

To show the position of these stars on the Hertzsprung-Russell(HR) diagram, we turned to the latest TESS catalog \citep{Stassun2019} to extract stellar parameters such as effective temperature, mass, surface gravity, stellar density, and luminosity for the HADS stars. These parameters are listed in Table \ref{HADS_para}.

Figure \ref{fig:HR_HADS} illustrates the HR diagram, which shows the newly identified HADS stars as colored dots. The theoretical boundaries of the instability strip from \citet{Murphy2019} and the evolutionary tracks computed using MESA v10108 with X = 0.71 and Z = 0.01 are also included \citep{Murphy2019}. The fundamental frequency values of each HADS star are marked on the diagram. By analyzing the distribution of these HADS stars in the diagram, we observe a consistent trend with the findings of \citet{Bowman2018,lv2022ApJ}, indicating that the main period of the star increases as it evolves on the main sequence. We classified these HADS stars into three groups based on surface gravity: $\log g \geq 4.0$ (Zero-age main sequence; ZAMS), $3.5 \leq \log g < 4.0$ (Mid-Age Main Sequence; MAMS), and $\log g < 3.5$ (Terminal-age main sequence; TAMS) \citep{Bowman2018,Hasanzadeh2021}. The evolutionary positions of these stars align with the corresponding log $g$. To further investigate these stars, we obtained the corresponding M$_{V}$ values from \citet{Zhou2023}, which are listed in Table \ref{HADS_para}. As shown in Figure \ref{fig:PLR}, the two stars that are further off in times signs are probably RR Lyrae stars, while the rest of the stars are consistent with the period-luminosity (P-L) relation established by \citet{2019MNRAS.486.4348Z}. The black dashed line represents the P-L relation for High-amplitude \dsct stars derived by \citet{2011AJ....142..110M}, while the blue dashed line represents the P-L relation obtained by \citet{2019MNRAS.486.4348Z}.

HADS stars are known for their intense pulsations, and they exhibit a period ratio (P1/P0) between the fundamental mode (P0) and the first overtone (P1) close to 0.77. When a HADS star pulsates in both the fundamental and first overtone radial modes, it is designated as a HADS(B) in the AAVSO International Variable Star Index. The P1/P0 ratio has been widely used as the primary parameter to distinguish between HADS(B) and simple HADS stars. Several studies in recent years have focused on finding a comprehensive model that can explain the relationship between P1/P0 and P0. Notable contributions include the works of \citet{Petersen1996, Poretti2005, Yang2022, Netzel2022}.

\citet{Furgoni2016} derived a best-fit equation for the P1/P0 and P0 relationship, given by P1/P0 = $–$0.084809 ($\pm$ 0.008298)P0 $+$ 0.782048 ($\pm$ 0.000995). In addition, \citet{Yang2021} compiled a catalog of double-mode and multi-mode HADS stars, consisting of 155 stars. Through statistical analysis of the 132 double-mode HADS stars pulsating in both P0 and P1, they found that most of these stars have a P0 value ranging from 0.050 to 0.175 days. The distribution of P0 exhibits a suggestion of a bimodal structure, with peaks at 0.06 and 0.09 days (see their Figure 1). \citet{Yang2021} derived an updated linear relation between P1/P0 and P0 as P1/P0 = $-$0.0940 ($\pm$ 0.0091)P0 $+$ 0.7835 ($\pm$ 0.0009).

\begin{figure}[h!]
\begin{center}
  \includegraphics[width=0.52\textwidth]{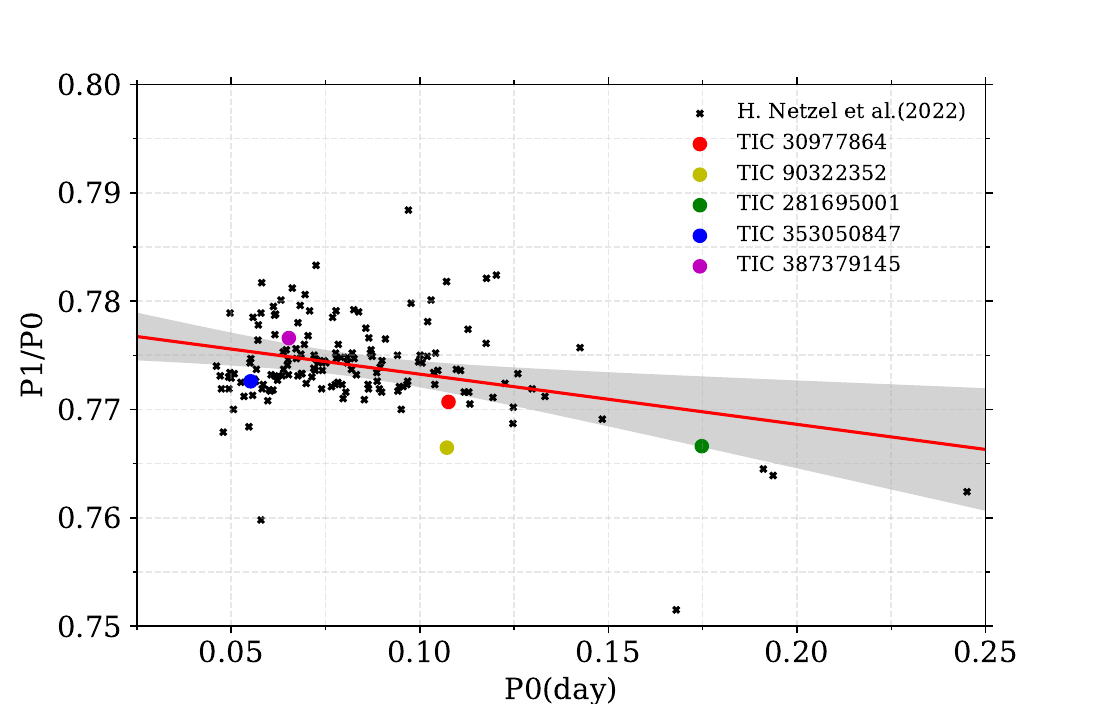}
  \caption{Petersen diagram of P1/P0 and P0. The different color dots indicate five HADS stars. The red line represents the linear fit of those data using emcee (a Markov chain Monte Carlo Ensemble sampler \citep{Foreman-Mackey2013}).}
    \label{fig:P0_P1}
\end{center}
\end{figure}

\begin{figure}[h!]
\begin{center}
  \includegraphics[width=0.48\textwidth]{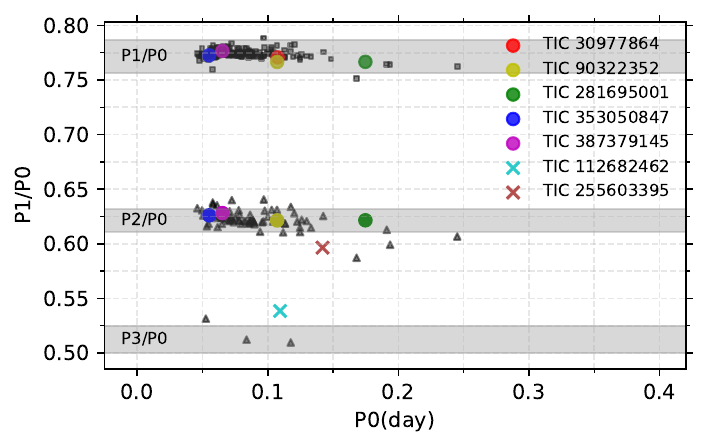}
  \caption{The Peterson diagram of the multi-mode pulsators, with black squares and black triangles from \citet{Netzel2022}. the circles of different colors represent the locations of the newly discovered HADS stars in this work, and the two times signs represent the two HADS stars without no first overtone detection. The gray shaded bands represent the ratio of the radial first overtone to the second overtone and the third overtone to the fundamental mode.}
    \label{fig:P0_P1_P2}
\end{center}
\end{figure}

\citet{Netzel2022} performed calculations using the MESA evolution code and the Warsaw pulsation code \citep{Dziembowski1977} to construct a grid of models. The grid spanned masses ranging from 0.8 $M_{\odot}$ to 2.5 $M_{\odot}$ and metallicities [Fe/H] ranging from -2.0 dex to +0.5 dex. Detailed parameter settings can be found in \citep{Netzel2022}. They obtained satisfactory fits for 176 HADS stars and reported their masses, metallicities, ages, luminosities, and effective temperatures. Among these stars, 16 candidates for SX Phoenicis stars. Using these 176 HADS stars, we derived a linear relation between P1/P0 and P0 as P1/P0 = $-$0.046($\pm$ 0.007)P0 $+$ 0.7778($\pm$ 0.0006), as shown in Figure \ref{fig:P0_P1}.

Figure \ref{fig:P0_P1} illustrates that the fundamental pulsation frequency of all HADS stars is below 26 day$^{-1}$, indicating that these HADS stars are likely evolved \dsct stars, with only a few belonging to the pre-main sequence phase. The scattering of P1/P0 values within the range of 0.05 to 0.10 days could potentially be attributed to stellar rotation. A study by \citet{suarez2006} investigated the effect of rotation on period ratios for double-mode pulsators and found that the differences in period ratios increase with rotational velocities for a given metallicity. They also discovered that the differences caused by rotation are comparable to those caused by metallicity up to 0.30 dex. Therefore, when accurately determining the mass and metallicity of a star, the influence of rotation should be taken into account. Additionally, near-degeneracy has been examined in relation to period ratios. \citet{suarez2007} investigated the influence of near-degeneracy on period ratios and found that it can significantly modify the period ratios, even for relatively slow rotators, by altering the oscillation frequencies through the coupling strength.

According to the metallicities of these stars derived by \citet{Netzel2022} using their models, the distribution of metallicity versus fundamental pulsation frequency P0 for these 176 HADS stars suggests a wide range of metallicities, from -2.0 dex to 0.5 dex, covering a period range of 0.05 to 0.20 days. It is possible that this broad distribution of metallicities contributes to the dispersion in the P1/P0 ratio observed in Figure \ref{fig:P0_P1}. This suggests that metallicities might have a more significant impact on the ratio of HADS stars compared to stellar rotation.

Figure \ref{fig:P0_P1_P2} presents the Peterson diagram of multi-mode pulsators, where the black squares represent the P1/P0 ratio and the black triangles represent the P2/P0 ratio from \citet{Netzel2022}. The colored circles indicate the positions of the newly discovered HADS stars in this study. The gray bands represent the range of ratios for the first overtone, second overtone, and third overtone with respect to the fundamental frequency. Two times signs are used to denote two HADS stars that do not exhibit the first overtone.

\section{Summary}
\label{sect:Summary}

In our study, we have identified seven new HADS stars. Among them, TIC 90322352 and TIC 387379145 appear to be pure pulsators without the excitation of additional non-radial modes. TIC 90322352 exhibits indications of being a new triple-mode HADS star, with a total of 26 detected frequencies. Similarly, TIC 387379145 also shows indications of being a new triple-mode HADS star, with a total of 40 detected frequencies. TIC 30977864 is classified as a double-mode HADS star, with a total of 11 detected frequencies. Interestingly, an additional frequency suggestive of non-radial pulsation mode was also detected in this star. TIC 112682462 is a double-mode HADS star, with the fundamental frequency and the third overtone detected among a total of 28 frequencies. The ratio of P3/P0 is measured as 0.538564, slightly larger than the theoretical ratio. Moreover, this star exhibits some frequencies in the low-frequency range, these frequencies may be a quintuple produced by the rotation, which adds to its intriguing nature. TIC 255603395 is another interesting case, showing indications of being a double-mode HADS star, with the fundamental frequency and the second overtone detected among a total of 31 frequencies. The ratio of P2/P0 is measured as 0.596479, slightly larger than the theoretical ratio, and non-radial frequencies were also detected. TIC 281695001 appears to be a new triple-mode HADS star, with a total of 49 frequencies detected. Notably, the amplitude of the first overtone is larger than the amplitude of the fundamental frequency, possibly due to variations in the amplitude of the fundamental frequency. Lastly, TIC 353050847 appears to be a triple-mode HADS star, with a total of 14 frequencies detected, and two frequencies possibly corresponding to non-radial modes.

In summary, our findings include four stars that potentially exhibit triple-mode HADS behavior, although the identification is not definitive due to the relatively small amplitude of the second overtone. Notably, TIC 112682462 and TIC 255603395 lack the first overtone and second overtone of the radial modes, respectively, making them particularly intriguing cases. The light curves of these two stars bear a resemblance to those of RR Lyrae stars. When comparing the distribution of TIC 112682462 and TIC 255603395 with several other HADS stars in Figure \ref{fig:HR_HADS} and Figure \ref{fig:PLR}, it becomes evident that the positions of these two stars differ from other HADS stars. However, due to the lack of accurate spectral parameters as support, we can only tentatively speculate that these two stars may be RR Lyrae stars.

Utilizing the 176 HADS stars from \citet{Netzel2022}, we derived a linear relation between P1/P0 and P0 as P1/P0 = $-$0.046($\pm$ 0.007)P0 $+$ 0.7778($\pm$ 0.0006). The distribution of metallicities among these stars is broad, ranging from -2.0 dex to 0.5 dex, for periods ranging from 0.05 to 0.20 days. This random distribution of metallicities may contribute to the dispersion in P1/P0. Our results indicate that an increase in [Fe/H] leads to a decrease in P1/P0 and P2/P0. To draw more precise conclusions, future spectroscopic observations of a larger sample of HADS stars are necessary. Such observations would not only provide accurate rotational velocities but also more precise determinations of metallicities.

\section*{Acknowledgments}

The work is supported by the National Key R\&D program of China for the Intergovernmental Scientific and Technological Innovation Cooperation Project under No. 2022YFE0126200. Chenglong lv would like to acknowledge the funding from the China Scholarship Council scholarship (No 202204910463). A.Hasanzadeh acknowledges support from the Science and Technology Facilities Council (STFC) grant No. ST/X000915/1. Some of the data presented in this paper were obtained from the Mikulski Archive for Space Telescopes (MAST). The Large Sky Area Multi-Object Fiber Spectroscopic Telescope (LAMOST) is a National Major Scientific Project built by the Chinese Academy of Sciences. Funding for the project has been provided by the National Development and Reform Commission. LAMOST is operated and managed by the National Astronomical Observatories, Chinese Academy of Sciences.

\bibliography{references}

@ARTICLE{Netzel2022,
       author = {{Netzel}, H. and {Smolec}, R.},
        title = "{Modelling of multimode radially pulsating high-amplitude {\ensuremath{\delta}} Scuti stars from the OGLE Galactic bulge sample}",
      journal = {\mnras},
         year = 2022,
        month = sep,
       volume = {515},
       number = {3},
        pages = {4574-4586},
          doi = {10.1093/mnras/stac1938},
archivePrefix = {arXiv},
       eprint = {2207.04210},
 primaryClass = {astro-ph.SR},
       adsurl = {https://ui.adsabs.harvard.edu/abs/2022MNRAS.515.4574N}
}

@ARTICLE{2003PASP..115.1023W,
       author = {{Walker}, Gordon and {Matthews}, Jaymie and {Kuschnig}, Rainer and {Johnson}, Ron and {Rucinski}, Slavek and {Pazder}, John and {Burley}, Gregory and {Walker}, Andrew and {Skaret}, Kristina and {Zee}, Robert and {Grocott}, Simon and {Carroll}, Kieran and {Sinclair}, Peter and {Sturgeon}, Don and {Harron}, John},
        title = "{The MOST Asteroseismology Mission: Ultraprecise Photometry from Space}",
      journal = {\pasp},
         year = 2003,
        month = sep,
       volume = {115},
       number = {811},
        pages = {1023-1035},
          doi = {10.1086/377358},
       adsurl = {https://ui.adsabs.harvard.edu/abs/2003PASP..115.1023W}
}

@ARTICLE{2009A&A...506..411A,
       author = {{Auvergne}, M. and {Bodin}, P. and {Boisnard}, L. and {Buey}, J. -T. and {Chaintreuil}, S. and {Epstein}, G. and {Jouret}, M. and {Lam-Trong}, T. and {Levacher}, P. and {Magnan}, A. and {Perez}, R. and {Plasson}, P. and {Plesseria}, J. and {Peter}, G. and {Steller}, M. and {Tiph{\`e}ne}, D. and {Baglin}, A. and {Agogu{\'e}}, P. and {Appourchaux}, T. and {Barbet}, D. and {Beaufort}, T. and {Bellenger}, R. and {Berlin}, R. and {Bernardi}, P. and {Blouin}, D. and {Boumier}, P. and {Bonneau}, F. and {Briet}, R. and {Butler}, B. and {Cautain}, R. and {Chiavassa}, F. and {Costes}, V. and {Cuvilho}, J. and {Cunha-Parro}, V. and {de Oliveira Fialho}, F. and {Decaudin}, M. and {Defise}, J. -M. and {Djalal}, S. and {Docclo}, A. and {Drummond}, R. and {Dupuis}, O. and {Exil}, G. and {Faur{\'e}}, C. and {Gaboriaud}, A. and {Gamet}, P. and {Gavalda}, P. and {Grolleau}, E. and {Gueguen}, L. and {Guivarc'h}, V. and {Guterman}, P. and {Hasiba}, J. and {Huntzinger}, G. and {Hustaix}, H. and {Imbert}, C. and {Jeanville}, G. and {Johlander}, B. and {Jorda}, L. and {Journoud}, P. and {Karioty}, F. and {Kerjean}, L. and {Lafond}, L. and {Lapeyrere}, V. and {Landiech}, P. and {Larqu{\'e}}, T. and {Laudet}, P. and {Le Merrer}, J. and {Leporati}, L. and {Leruyet}, B. and {Levieuge}, B. and {Llebaria}, A. and {Martin}, L. and {Mazy}, E. and {Mesnager}, J. -M. and {Michel}, J. -P. and {Moalic}, J. -P. and {Monjoin}, W. and {Naudet}, D. and {Neukirchner}, S. and {Nguyen-Kim}, K. and {Ollivier}, M. and {Orcesi}, J. -L. and {Ottacher}, H. and {Oulali}, A. and {Parisot}, J. and {Perruchot}, S. and {Piacentino}, A. and {Pinheiro da Silva}, L. and {Platzer}, J. and {Pontet}, B. and {Pradines}, A. and {Quentin}, C. and {Rohbeck}, U. and {Rolland}, G. and {Rollenhagen}, F. and {Romagnan}, R. and {Russ}, N. and {Samadi}, R. and {Schmidt}, R. and {Schwartz}, N. and {Sebbag}, I. and {Smit}, H. and {Sunter}, W. and {Tello}, M. and {Toulouse}, P. and {Ulmer}, B. and {Vandermarcq}, O. and {Vergnault}, E. and {Wallner}, R. and {Waultier}, G. and {Zanatta}, P.},
        title = "{The CoRoT satellite in flight: description and performance}",
      journal = {\aap},
         year = 2009,
        month = oct,
       volume = {506},
       number = {1},
        pages = {411-424},
          doi = {10.1051/0004-6361/200810860},
archivePrefix = {arXiv},
       eprint = {0901.2206},
 primaryClass = {astro-ph.SR},
       adsurl = {https://ui.adsabs.harvard.edu/abs/2009A&A...506..411A}
}

@ARTICLE{2010PASP..122..131G,
       author = {{Gilliland}, Ronald L. and {Brown}, Timothy M. and {Christensen-Dalsgaard}, J{\o}rgen and {Kjeldsen}, Hans and {Aerts}, Conny and {Appourchaux}, Thierry and {Basu}, Sarbani and {Bedding}, Timothy R. and {Chaplin}, William J. and {Cunha}, Margarida S. and {De Cat}, Peter and {De Ridder}, Joris and {Guzik}, Joyce A. and {Handler}, Gerald and {Kawaler}, Steven and {Kiss}, L{\'a}szl{\'o} and {Kolenberg}, Katrien and {Kurtz}, Donald W. and {Metcalfe}, Travis S. and {Monteiro}, Mario J.~P.~F.~G. and {Szab{\'o}}, Robert and {Arentoft}, Torben and {Balona}, Luis and {Debosscher}, Jonas and {Elsworth}, Yvonne P. and {Quirion}, Pierre-Olivier and {Stello}, Dennis and {Su{\'a}rez}, Juan Carlos and {Borucki}, William J. and {Jenkins}, Jon M. and {Koch}, David and {Kondo}, Yoji and {Latham}, David W. and {Rowe}, Jason F. and {Steffen}, Jason H.},
        title = "{Kepler Asteroseismology Program: Introduction and First Results}",
      journal = {\pasp},
         year = 2010,
        month = feb,
       volume = {122},
       number = {888},
        pages = {131},
          doi = {10.1086/650399},
archivePrefix = {arXiv},
       eprint = {1001.0139},
 primaryClass = {astro-ph.SR},
       adsurl = {https://ui.adsabs.harvard.edu/abs/2010PASP..122..131G}
}

@ARTICLE{2010ApJ...713L..79K,
       author = {{Koch}, David G. and {Borucki}, William J. and {Basri}, Gibor and {Batalha}, Natalie M. and {Brown}, Timothy M. and {Caldwell}, Douglas and {Christensen-Dalsgaard}, J{\o}rgen and {Cochran}, William D. and {DeVore}, Edna and {Dunham}, Edward W. and {Gautier}, Thomas N., III and {Geary}, John C. and {Gilliland}, Ronald L. and {Gould}, Alan and {Jenkins}, Jon and {Kondo}, Yoji and {Latham}, David W. and {Lissauer}, Jack J. and {Marcy}, Geoffrey and {Monet}, David and {Sasselov}, Dimitar and {Boss}, Alan and {Brownlee}, Donald and {Caldwell}, John and {Dupree}, Andrea K. and {Howell}, Steve B. and {Kjeldsen}, Hans and {Meibom}, S{\o}ren and {Morrison}, David and {Owen}, Tobias and {Reitsema}, Harold and {Tarter}, Jill and {Bryson}, Stephen T. and {Dotson}, Jessie L. and {Gazis}, Paul and {Haas}, Michael R. and {Kolodziejczak}, Jeffrey and {Rowe}, Jason F. and {Van Cleve}, Jeffrey E. and {Allen}, Christopher and {Chandrasekaran}, Hema and {Clarke}, Bruce D. and {Li}, Jie and {Quintana}, Elisa V. and {Tenenbaum}, Peter and {Twicken}, Joseph D. and {Wu}, Hayley},
        title = "{Kepler Mission Design, Realized Photometric Performance, and Early Science}",
      journal = {\apjl},
         year = 2010,
        month = apr,
       volume = {713},
       number = {2},
        pages = {L79-L86},
          doi = {10.1088/2041-8205/713/2/L79},
archivePrefix = {arXiv},
       eprint = {1001.0268},
 primaryClass = {astro-ph.EP},
       adsurl = {https://ui.adsabs.harvard.edu/abs/2010ApJ...713L..79K}
}

@ARTICLE{1994ARA&A..32...37B,
       author = {{Brown}, Timothy M. and {Gilliland}, Ronald L.},
        title = "{Asteroseismology}",
      journal = {\araa},
         year = 1994,
        month = jan,
       volume = {32},
        pages = {37-82},
          doi = {10.1146/annurev.aa.32.090194.000345},
       adsurl = {https://ui.adsabs.harvard.edu/abs/1994ARA&A..32...37B}
}

@ARTICLE{Dupret2004,
       author = {{Dupret}, M. -A. and {Thoul}, A. and {Scuflaire}, R. and {Daszy{\'n}ska-Daszkiewicz}, J. and {Aerts}, C. and {Bourge}, P. -O. and {Waelkens}, C. and {Noels}, A.},
        title = "{Asteroseismology of the {\ensuremath{\beta}} Cep star HD 129929. II. Seismic constraints on core overshooting, internal rotation  and stellar parameters}",
      journal = {\aap},
         year = 2004,
        month = feb,
       volume = {415},
        pages = {251-257},
          doi = {10.1051/0004-6361:20034143},
       adsurl = {https://ui.adsabs.harvard.edu/abs/2004A&A...415..251D}
}

@BOOK{ASTERO,
       author = {{Aerts}, Conny and {Christensen-Dalsgaard}, J{\o}rgen and {Kurtz}, Donald W.},
        title = "{Asteroseismology}",
         year = 2010,
          doi = {10.1007/978-1-4020-5803-5},
       adsurl = {https://ui.adsabs.harvard.edu/abs/2010aste.book.....A}
}

@BOOK{2015pust.book.....C,
       author = {{Catelan}, M. and {Smith}, H.~A.},
        title = "{Pulsating Stars}",
         year = 2015,
       adsurl = {https://ui.adsabs.harvard.edu/abs/2015pust.book.....C}
}

@ARTICLE{2019LRSP...16....4G,
       author = {{Garc{\'\i}a}, Rafael A. and {Ballot}, J{\'e}r{\^o}me},
        title = "{Asteroseismology of solar-type stars}",
      journal = {Living Reviews in Solar Physics},
         year = 2019,
        month = sep,
       volume = {16},
       number = {1},
          eid = {4},
        pages = {4},
          doi = {10.1007/s41116-019-0020-1},
archivePrefix = {arXiv},
       eprint = {1906.12262},
 primaryClass = {astro-ph.SR},
       adsurl = {https://ui.adsabs.harvard.edu/abs/2019LRSP...16....4G}
}

@ARTICLE{Daszy2022,
       author = {{Daszy{\'n}ska-Daszkiewicz}, J. and {Walczak}, P. and {Pamyatnykh}, A.~A. and {Szewczuk}, W.},
        title = "{Asteroseismology of the double-radial mode {\ensuremath{\delta}} Scuti star BP Pegasi}",
      journal = {\mnras},
         year = 2022,
        month = may,
       volume = {512},
       number = {3},
        pages = {3551-3565},
          doi = {10.1093/mnras/stac646},
archivePrefix = {arXiv},
       eprint = {2203.02722},
 primaryClass = {astro-ph.SR},
       adsurl = {https://ui.adsabs.harvard.edu/abs/2022MNRAS.512.3551D}
}

@ARTICLE{2016ApJ...830..138C,
       author = {{Campante}, T.~L. and {Schofield}, M. and {Kuszlewicz}, J.~S. and {Bouma}, L. and {Chaplin}, W.~J. and {Huber}, D. and {Christensen-Dalsgaard}, J. and {Kjeldsen}, H. and {Bossini}, D. and {North}, T.~S.~H. and {Appourchaux}, T. and {Latham}, D.~W. and {Pepper}, J. and {Ricker}, G.~R. and {Stassun}, K.~G. and {Vanderspek}, R. and {Winn}, J.~N.},
        title = "{The Asteroseismic Potential of TESS: Exoplanet-host Stars}",
      journal = {\apj},
         year = 2016,
        month = oct,
       volume = {830},
       number = {2},
          eid = {138},
        pages = {138},
          doi = {10.3847/0004-637X/830/2/138},
archivePrefix = {arXiv},
       eprint = {1608.01138},
 primaryClass = {astro-ph.SR},
       adsurl = {https://ui.adsabs.harvard.edu/abs/2016ApJ...830..138C}
}

@ARTICLE{2019AJ....157..245H,
       author = {{Huber}, Daniel and {Chaplin}, William J. and {Chontos}, Ashley and {Kjeldsen}, Hans and {Christensen-Dalsgaard}, J{\o}rgen and {Bedding}, Timothy R. and {Ball}, Warrick and {Brahm}, Rafael and {Espinoza}, Nestor and {Henning}, Thomas and {Jord{\'a}n}, Andr{\'e}s and {Sarkis}, Paula and {Knudstrup}, Emil and {Albrecht}, Simon and {Grundahl}, Frank and {Fredslund Andersen}, Mads and {Pall{\'e}}, Pere L. and {Crossfield}, Ian and {Fulton}, Benjamin and {Howard}, Andrew W. and {Isaacson}, Howard T. and {Weiss}, Lauren M. and {Handberg}, Rasmus and {Lund}, Mikkel N. and {Serenelli}, Aldo M. and {R{\o}rsted Mosumgaard}, Jakob and {Stokholm}, Amalie and {Bieryla}, Allyson and {Buchhave}, Lars A. and {Latham}, David W. and {Quinn}, Samuel N. and {Gaidos}, Eric and {Hirano}, Teruyuki and {Ricker}, George R. and {Vanderspek}, Roland K. and {Seager}, Sara and {Jenkins}, Jon M. and {Winn}, Joshua N. and {Antia}, H.~M. and {Appourchaux}, Thierry and {Basu}, Sarbani and {Bell}, Keaton J. and {Benomar}, Othman and {Bonanno}, Alfio and {Buzasi}, Derek L. and {Campante}, Tiago L. and {{\c{C}}elik Orhan}, Z. and {Corsaro}, Enrico and {Cunha}, Margarida S. and {Davies}, Guy R. and {Deheuvels}, Sebastien and {Grunblatt}, Samuel K. and {Hasanzadeh}, Amir and {Di Mauro}, Maria Pia and {Garc{\'\i}a}, Rafael A. and {Gaulme}, Patrick and {Girardi}, L{\'e}o and {Guzik}, Joyce A. and {Hon}, Marc and {Jiang}, Chen and {Kallinger}, Thomas and {Kawaler}, Steven D. and {Kuszlewicz}, James S. and {Lebreton}, Yveline and {Li}, Tanda and {Lucas}, Miles and {Lundkvist}, Mia S. and {Mann}, Andrew W. and {Mathis}, St{\'e}phane and {Mathur}, Savita and {Mazumdar}, Anwesh and {Metcalfe}, Travis S. and {Miglio}, Andrea and {Monteiro}, M{\'a}rio J.~P.~F.~G. and {Mosser}, Benoit and {Noll}, Anthony and {Nsamba}, Benard and {Ong}, Jia Mian Joel and {{\"O}rtel}, S. and {Pereira}, Filipe and {Ranadive}, Pritesh and {R{\'e}gulo}, Clara and {Rodrigues}, Tha{\'\i}se S. and {Roxburgh}, Ian W. and {Silva Aguirre}, Victor and {Smalley}, Barry and {Schofield}, Mathew and {Sousa}, S{\'e}rgio G. and {Stassun}, Keivan G. and {Stello}, Dennis and {Tayar}, Jamie and {White}, Timothy R. and {Verma}, Kuldeep and {Vrard}, Mathieu and {Y{\i}ld{\i}z}, M. and {Baker}, David and {Bazot}, Micha{\"e}l and {Beichmann}, Charles and {Bergmann}, Christoph and {Bugnet}, Lisa and {Cale}, Bryson and {Carlino}, Roberto and {Cartwright}, Scott M. and {Christiansen}, Jessie L. and {Ciardi}, David R. and {Creevey}, Orlagh and {Dittmann}, Jason A. and {Do Nascimento}, Jose-Dias, Jr. and {Van Eylen}, Vincent and {F{\"u}r{\'e}sz}, Gabor and {Gagn{\'e}}, Jonathan and {Gao}, Peter and {Gazeas}, Kosmas and {Giddens}, Frank and {Hall}, Oliver J. and {Hekker}, Saskia and {Ireland}, Michael J. and {Latouf}, Natasha and {LeBrun}, Danny and {Levine}, Alan M. and {Matzko}, William and {Natinsky}, Eva and {Page}, Emma and {Plavchan}, Peter and {Mansouri-Samani}, Masoud and {McCauliff}, Sean and {Mullally}, Susan E. and {Orenstein}, Brendan and {Garcia Soto}, Aylin and {Paegert}, Martin and {van Saders}, Jennifer L. and {Schnaible}, Chloe and {Soderblom}, David R. and {Szab{\'o}}, R{\'o}bert and {Tanner}, Angelle and {Tinney}, C.~G. and {Teske}, Johanna and {Thomas}, Alexandra and {Trampedach}, Regner and {Wright}, Duncan and {Yuan}, Thomas T. and {Zohrabi}, Farzaneh},
        title = "{A Hot Saturn Orbiting an Oscillating Late Subgiant Discovered by TESS}",
      journal = {\aj},
         year = 2019,
        month = jun,
       volume = {157},
       number = {6},
          eid = {245},
        pages = {245},
          doi = {10.3847/1538-3881/ab1488},
archivePrefix = {arXiv},
       eprint = {1901.01643},
 primaryClass = {astro-ph.EP},
       adsurl = {https://ui.adsabs.harvard.edu/abs/2019AJ....157..245H}
}

@INPROCEEDINGS{McNamara2000,
       author = {{McNamara}, D.~H.},
        title = "{The High-Amplitude {\ensuremath{\delta}} Scuti Stars}",
    booktitle = {Delta Scuti and Related Stars},
         year = 2000,
       editor = {{Breger}, Michel and {Montgomery}, Michael},
       series = {Astronomical Society of the Pacific Conference Series},
       volume = {210},
        month = jan,
        pages = {373},
       adsurl = {https://ui.adsabs.harvard.edu/abs/2000ASPC..210..373M}
}

@ARTICLE{Poretti2011,
       author = {{Poretti}, E. and {Rainer}, M. and {Weiss}, W.~W. and {Bogn{\'a}r}, Zs. and {Moya}, A. and {Niemczura}, E. and {Su{\'a}rez}, J.~C. and {Auvergne}, M. and {Baglin}, A. and {Baudin}, F. and {Benk{\H{o}}}, J.~M. and {Debosscher}, J. and {Garrido}, R. and {Mantegazza}, L. and {Papar{\'o}}, M.},
        title = "{Monitoring a high-amplitude {\ensuremath{\delta}} Scuti star for 152 days: discovery of 12 additional modes and modulation effects in the light curve of CoRoT 101155310}",
      journal = {\aap},
         year = 2011,
        month = apr,
       volume = {528},
          eid = {A147},
        pages = {A147},
          doi = {10.1051/0004-6361/201016045},
archivePrefix = {arXiv},
       eprint = {1102.3085},
 primaryClass = {astro-ph.SR},
       adsurl = {https://ui.adsabs.harvard.edu/abs/2011A&A...528A.147P}
}

@ARTICLE{Bowman2021,
       author = {{Bowman}, D.~M. and {Hermans}, J. and {Daszy{\'n}ska-Daszkiewicz}, J. and {Holdsworth}, D.~L. and {Tkachenko}, A. and {Murphy}, S.~J. and {Smalley}, B. and {Kurtz}, D.~W.},
        title = "{KIC 5950759: a high-amplitude {\ensuremath{\delta}} Sct star with amplitude and frequency modulation near the terminal age main sequence}",
      journal = {\mnras},
         year = 2021,
        month = jul,
       volume = {504},
       number = {3},
        pages = {4039-4053},
          doi = {10.1093/mnras/stab1124},
archivePrefix = {arXiv},
       eprint = {2104.09279},
 primaryClass = {astro-ph.SR},
       adsurl = {https://ui.adsabs.harvard.edu/abs/2021MNRAS.504.4039B}
}

@ARTICLE{lv2022AJ,
       author = {{Lv}, Chenglong and {Esamdin}, Ali and {Pascual-Granado}, J. and {Hern{\'a}ndez}, A. Garc{\'\i}a and {Hasanzadeh}, A.},
        title = "{Asteroseismology of a Double-mode High-amplitude {\ensuremath{\delta}} Scuti Star TIC 448892817}",
      journal = {\aj},
         year = 2022,
        month = nov,
       volume = {164},
       number = {5},
          eid = {218},
        pages = {218},
          doi = {10.3847/1538-3881/ac9473},
       adsurl = {https://ui.adsabs.harvard.edu/abs/2022AJ....164..218L}
}

@ARTICLE{Yang2022,
       author = {{Yang}, Tao-Zhi and {Zuo}, Zhao-Yu and {Sun}, Xiao-Ya and {Tang}, Rui-Xuan and {Esamdin}, Ali},
        title = "{KIC 2857323: A Double-mode High-amplitude {\ensuremath{\delta}} Scuti Star with Amplitude Decline of the First Overtone Mode}",
      journal = {\apj},
         year = 2022,
        month = sep,
       volume = {936},
       number = {1},
          eid = {48},
        pages = {48},
          doi = {10.3847/1538-4357/ac86c9},
       adsurl = {https://ui.adsabs.harvard.edu/abs/2022ApJ...936...48Y}
}

@ARTICLE{Lares2020,
       author = {{Lares-Martiz}, M. and {Garrido}, R. and {Pascual-Granado}, J.},
        title = "{Self-consistent method to extract non-linearities from pulsating star light curves - I. Combination frequencies}",
      journal = {\mnras},
         year = 2020,
        month = oct,
       volume = {498},
       number = {1},
        pages = {1194-1204},
          doi = {10.1093/mnras/staa2256},
archivePrefix = {arXiv},
       eprint = {2005.10901},
 primaryClass = {astro-ph.SR},
       adsurl = {https://ui.adsabs.harvard.edu/abs/2020MNRAS.498.1194L}
}

@ARTICLE{Petersen1973,
       author = {{Petersen}, J.~O.},
        title = "{Masses of double mode cepheid variables determined by analysis of period ratios.}",
      journal = {\aap},
         year = 1973,
        month = aug,
       volume = {27},
        pages = {89},
       adsurl = {https://ui.adsabs.harvard.edu/abs/1973A&A....27...89P}
}

@ARTICLE{lv2022ApJ,
       author = {{Lv}, Chenglong and {Esamdin}, Ali and {Pascual-Granado}, J. and {Yang}, Taozhi and {Shen}, Dongxiang},
        title = "{Frequency Analysis of KIC 1573174: Shedding Light on the Nature of HADS Stars}",
      journal = {\apj},
         year = 2022,
        month = jun,
       volume = {932},
       number = {1},
          eid = {42},
        pages = {42},
          doi = {10.3847/1538-4357/ac69d9},
archivePrefix = {arXiv},
       eprint = {2205.00571},
 primaryClass = {astro-ph.SR},
       adsurl = {https://ui.adsabs.harvard.edu/abs/2022ApJ...932...42L}
}

@ARTICLE{Daszy2020,
       author = {{Daszy{\'n}ska-Daszkiewicz}, J. and {Pamyatnykh}, A.~A. and {Walczak}, P. and {Szewczuk}, W.},
        title = "{Seismic analysis of the double-mode radial pulsator SX Phoenicis}",
      journal = {\mnras},
         year = 2020,
        month = dec,
       volume = {499},
       number = {2},
        pages = {3034-3045},
          doi = {10.1093/mnras/staa3056},
archivePrefix = {arXiv},
       eprint = {2009.14065},
 primaryClass = {astro-ph.SR},
       adsurl = {https://ui.adsabs.harvard.edu/abs/2020MNRAS.499.3034D}
}

@BOOK{Bowman2017,
       author = {{Bowman}, Dominic M.},
        title = "{Amplitude Modulation of Pulsation Modes in Delta Scuti Stars}",
        journal = {Springer Theses series. ISBN 978-3-319-66649-5},
         year = 2017,
          doi = {10.1007/978-3-319-66649-5},
       adsurl = {https://ui.adsabs.harvard.edu/abs/2017ampm.book.....B}
}

@ARTICLE{Mow2016,
       author = {{Mow}, Benjamin and {Reinhart}, Erik and {Nhim}, Samantha and {Watkins}, Richard},
        title = "{Rapid Evolution of GSC 03144-595, a New Triple-mode Radially Pulsating High-amplitude {\ensuremath{\delta}} Scuti}",
      journal = {\aj},
         year = 2016,
        month = jul,
       volume = {152},
       number = {1},
          eid = {17},
        pages = {17},
          doi = {10.3847/0004-6256/152/1/17},
       adsurl = {https://ui.adsabs.harvard.edu/abs/2016AJ....152...17M}
}

@INPROCEEDINGS{Jenkins2016,
       author = {{Jenkins}, Jon M. and {Twicken}, Joseph D. and {McCauliff}, Sean and {Campbell}, Jennifer and {Sanderfer}, Dwight and {Lung}, David and {Mansouri-Samani}, Masoud and {Girouard}, Forrest and {Tenenbaum}, Peter and {Klaus}, Todd and {Smith}, Jeffrey C. and {Caldwell}, Douglas A. and {Chacon}, A.~D. and {Henze}, Christopher and {Heiges}, Cory and {Latham}, David W. and {Morgan}, Edward and {Swade}, Daryl and {Rinehart}, Stephen and {Vanderspek}, Roland},
        title = "{The TESS science processing operations center}",
    booktitle = {Software and Cyberinfrastructure for Astronomy IV},
         year = 2016,
       editor = {{Chiozzi}, Gianluca and {Guzman}, Juan C.},
       series = {Society of Photo-Optical Instrumentation Engineers (SPIE) Conference Series},
       volume = {9913},
        month = aug,
          eid = {99133E},
        pages = {99133E},
          doi = {10.1117/12.2233418},
       adsurl = {https://ui.adsabs.harvard.edu/abs/2016SPIE.9913E..3EJ}
}

@ARTICLE{Zhou2023,
       author = {{Zhou}, A. -Y.},
        title = "{A Catalog of 59 Thousand $\delta$ Scuti Stars and Dozen Discoveries of New Variables with TESS Data}",
      journal = {arXiv e-prints},
         year = 2023,
        month = jan,
          eid = {arXiv:2301.08355},
        pages = {arXiv:2301.08355},
          doi = {10.48550/arXiv.2301.08355},
archivePrefix = {arXiv},
       eprint = {2301.08355},
 primaryClass = {astro-ph.SR},
       adsurl = {https://ui.adsabs.harvard.edu/abs/2023arXiv230108355Z}
}

@ARTICLE{Cui2012,
       author = {{Cui}, Xiang-Qun and {Zhao}, Yong-Heng and {Chu}, Yao-Quan and {Li}, Guo-Ping and {Li}, Qi and {Zhang}, Li-Ping and {Su}, Hong-Jun and {Yao}, Zheng-Qiu and {Wang}, Ya-Nan and {Xing}, Xiao-Zheng and {Li}, Xin-Nan and {Zhu}, Yong-Tian and {Wang}, Gang and {Gu}, Bo-Zhong and {Luo}, A. -Li and {Xu}, Xin-Qi and {Zhang}, Zhen-Chao and {Liu}, Gen-Rong and {Zhang}, Hao-Tong and {Yang}, De-Hua and {Cao}, Shu-Yun and {Chen}, Hai-Yuan and {Chen}, Jian-Jun and {Chen}, Kun-Xin and {Chen}, Ying and {Chu}, Jia-Ru and {Feng}, Lei and {Gong}, Xue-Fei and {Hou}, Yong-Hui and {Hu}, Hong-Zhuan and {Hu}, Ning-Sheng and {Hu}, Zhong-Wen and {Jia}, Lei and {Jiang}, Fang-Hua and {Jiang}, Xiang and {Jiang}, Zi-Bo and {Jin}, Ge and {Li}, Ai-Hua and {Li}, Yan and {Li}, Ye-Ping and {Liu}, Guan-Qun and {Liu}, Zhi-Gang and {Lu}, Wen-Zhi and {Mao}, Yin-Dun and {Men}, Li and {Qi}, Yong-Jun and {Qi}, Zhao-Xiang and {Shi}, Huo-Ming and {Tang}, Zheng-Hong and {Tao}, Qing-Sheng and {Wang}, Da-Qi and {Wang}, Dan and {Wang}, Guo-Min and {Wang}, Hai and {Wang}, Jia-Ning and {Wang}, Jian and {Wang}, Jian-Ling and {Wang}, Jian-Ping and {Wang}, Lei and {Wang}, Shu-Qing and {Wang}, You and {Wang}, Yue-Fei and {Xu}, Ling-Zhe and {Xu}, Yan and {Yang}, Shi-Hai and {Yu}, Yong and {Yuan}, Hui and {Yuan}, Xiang-Yan and {Zhai}, Chao and {Zhang}, Jing and {Zhang}, Yan-Xia and {Zhang}, Yong and {Zhao}, Ming and {Zhou}, Fang and {Zhou}, Guo-Hua and {Zhu}, Jie and {Zou}, Si-Cheng},
        title = "{The Large Sky Area Multi-Object Fiber Spectroscopic Telescope (LAMOST)}",
      journal = {Research in Astronomy and Astrophysics},
         year = 2012,
        month = sep,
       volume = {12},
       number = {9},
        pages = {1197-1242},
          doi = {10.1088/1674-4527/12/9/003},
       adsurl = {https://ui.adsabs.harvard.edu/abs/2012RAA....12.1197C}
}

@INPROCEEDINGS{Twicken2010,
       author = {{Twicken}, Joseph D. and {Chandrasekaran}, Hema and {Jenkins}, Jon M. and {Gunter}, Jay P. and {Girouard}, Forrest and {Klaus}, Todd C.},
        title = "{Presearch data conditioning in the Kepler Science Operations Center pipeline}",
    booktitle = {Software and Cyberinfrastructure for Astronomy},
         year = 2010,
       editor = {{Radziwill}, Nicole M. and {Bridger}, Alan},
       series = {Society of Photo-Optical Instrumentation Engineers (SPIE) Conference Series},
       volume = {7740},
        month = jul,
          eid = {77401U},
        pages = {77401U},
          doi = {10.1117/12.856798},
       adsurl = {https://ui.adsabs.harvard.edu/abs/2010SPIE.7740E..1UT}
}

@ARTICLE{Loumos1978,
       author = {{Loumos}, G.~L. and {Deeming}, T.~J.},
        title = "{Spurious Results from Fourier Analysis of Data with Closely Spaced Frequencies}",
      journal = {\apss},
         year = 1978,
        month = jul,
       volume = {56},
       number = {2},
        pages = {285-291},
          doi = {10.1007/BF01879560},
       adsurl = {https://ui.adsabs.harvard.edu/abs/1978Ap&SS..56..285L}
}

@ARTICLE{Charpinet2010,
       author = {{Charpinet}, S. and {Green}, E.~M. and {Baglin}, A. and {Van Grootel}, V. and {Fontaine}, G. and {Vauclair}, G. and {Chaintreuil}, S. and {Weiss}, W.~W. and {Michel}, E. and {Auvergne}, M. and {Catala}, C. and {Samadi}, R. and {Baudin}, F.},
        title = "{CoRoT opens a new era in hot B subdwarf asteroseismology. Detection of multiple g-mode oscillations in KPD 0629-0016}",
      journal = {\aap},
         year = 2010,
        month = jun,
       volume = {516},
          eid = {L6},
        pages = {L6},
          doi = {10.1051/0004-6361/201014789},
       adsurl = {https://ui.adsabs.harvard.edu/abs/2010A&A...516L...6C}
}

@ARTICLE{Zong2016,
       author = {{Zong}, W. and {Charpinet}, S. and {Vauclair}, G.},
        title = "{Signatures of nonlinear mode interactions in the pulsating hot B subdwarf star KIC 10139564}",
      journal = {\aap},
         year = 2016,
        month = oct,
       volume = {594},
          eid = {A46},
        pages = {A46},
          doi = {10.1051/0004-6361/201629132},
archivePrefix = {arXiv},
       eprint = {1607.06621},
 primaryClass = {astro-ph.SR},
       adsurl = {https://ui.adsabs.harvard.edu/abs/2016A&A...594A..46Z}
}

@ARTICLE{Baran2015,
       author = {{Baran}, A.~S. and {Koen}, C. and {Pokrzywka}, B.},
        title = "{A detection threshold in the amplitude spectra calculated from Kepler data obtained during K2 mission.}",
      journal = {\mnras},
         year = 2015,
        month = mar,
       volume = {448},
        pages = {L16-L19},
          doi = {10.1093/mnrasl/slu194},
archivePrefix = {arXiv},
       eprint = {1412.2753},
 primaryClass = {astro-ph.SR},
       adsurl = {https://ui.adsabs.harvard.edu/abs/2015MNRAS.448L..16B}
}

@ARTICLE{1999DSSN...13...28M,
       author = {{Montgomery}, M.~H. and {O'Donoghue}, D.},
        title = "{A derivation of the errors for least squares fitting to time series data}",
      journal = {Delta Scuti Star Newsletter},
         year = 1999,
        month = jul,
       volume = {13},
        pages = {28},
       adsurl = {https://ui.adsabs.harvard.edu/abs/1999DSSN...13...28M}
}

@ARTICLE{Tian2014,
       author = {{Tian}, Z.~J. and {Bi}, S.~L. and {Yang}, W.~M. and {Chen}, Y.~Q. and {Liu}, Z.~E. and {Liu}, K. and {Li}, T.~D. and {Ge}, Z.~S. and {Yu}, J.},
        title = "{Asteroseismic analysis of solar-like star KIC 6225718: constraints on stellar parameters and core overshooting}",
      journal = {\mnras},
         year = 2014,
        month = dec,
       volume = {445},
       number = {3},
        pages = {2999-3008},
          doi = {10.1093/mnras/stu1768},
       adsurl = {https://ui.adsabs.harvard.edu/abs/2014MNRAS.445.2999T}
}

@ARTICLE{Kern2018,
       author = {{Kern}, J.~W. and {Reed}, M.~D. and {Baran}, A.~S. and {Telting}, J.~H. and {{\O}stensen}, R.~H.},
        title = "{Asteroseismic analysis of the pulsating subdwarf B star KIC 11558725: an sdB+WD system with divergent frequency multiplets and mode trapping observed by Kepler}",
      journal = {\mnras},
         year = 2018,
        month = mar,
       volume = {474},
       number = {4},
        pages = {4709-4716},
          doi = {10.1093/mnras/stx2893},
       adsurl = {https://ui.adsabs.harvard.edu/abs/2018MNRAS.474.4709K}
}

@ARTICLE{Guthnick1933,
       author = {{Guthnick}, P. and {Prager}, R.},
        title = "{Benennung von ver{\"a}nderlichen Sternen}",
      journal = {Astronomische Nachrichten},
         year = 1933,
        month = jul,
       volume = {249},
        pages = {253},
          doi = {10.1002/asna.19332491502},
       adsurl = {https://ui.adsabs.harvard.edu/abs/1933AN....249..253G}
}

@ARTICLE{Stellingwerf1979,
       author = {{Stellingwerf}, R.~F.},
        title = "{Pulsation in the lower Cepheid strip. I. Linear survey.}",
      journal = {\apj},
         year = 1979,
        month = feb,
       volume = {227},
        pages = {935-942},
          doi = {10.1086/156802},
       adsurl = {https://ui.adsabs.harvard.edu/abs/1979ApJ...227..935S}
}

@INPROCEEDINGS{Breger2000,
       author = {{Breger}, M.},
        title = "{{\ensuremath{\delta}} Scuti stars (Review)}",
    booktitle = {Delta Scuti and Related Stars},
         year = 2000,
       editor = {{Breger}, Michel and {Montgomery}, Michael},
       series = {Astronomical Society of the Pacific Conference Series},
       volume = {210},
        month = jan,
        pages = {3},
       adsurl = {https://ui.adsabs.harvard.edu/abs/2000ASPC..210....3B}
}

@ARTICLE{Kurtz2015,
       author = {{Kurtz}, Donald W. and {Shibahashi}, Hiromoto and {Murphy}, Simon J. and {Bedding}, Timothy R. and {Bowman}, Dominic M.},
        title = "{A unifying explanation of complex frequency spectra of {\ensuremath{\gamma}} Dor, SPB and Be stars: combination frequencies and highly non-sinusoidal light curves}",
      journal = {\mnras},
         year = 2015,
        month = jul,
       volume = {450},
       number = {3},
        pages = {3015-3029},
          doi = {10.1093/mnras/stv868},
archivePrefix = {arXiv},
       eprint = {1504.04245},
 primaryClass = {astro-ph.SR},
       adsurl = {https://ui.adsabs.harvard.edu/abs/2015MNRAS.450.3015K}
}

@ARTICLE{Barac2022,
       author = {{Barac}, Natascha and {Bedding}, Timothy R. and {Murphy}, Simon J. and {Hey}, Daniel R.},
        title = "{Revisiting bright {\ensuremath{\delta}} Scuti stars and their period-luminosity relation with TESS and Gaia DR3}",
      journal = {\mnras},
         year = 2022,
        month = oct,
       volume = {516},
       number = {2},
        pages = {2080-2094},
          doi = {10.1093/mnras/stac2132},
archivePrefix = {arXiv},
       eprint = {2207.00343},
 primaryClass = {astro-ph.SR},
       adsurl = {https://ui.adsabs.harvard.edu/abs/2022MNRAS.516.2080B}
}

@ARTICLE{Kotov1987,
       author = {{Kotov}, V.~A.},
        title = "{The oscillation period of Delta Scuti stars close to 160 minutes.}",
      journal = {Izvestiya Ordena Trudovogo Krasnogo Znameni Krymskoj Astrofizicheskoj Observatorii},
         year = 1987,
        month = jan,
       volume = {76},
        pages = {10-20},
       adsurl = {https://ui.adsabs.harvard.edu/abs/1987IzKry..76...10K}
}

@ARTICLE{Clementini2023,
       author = {{Clementini}, G. and {Ripepi}, V. and {Garofalo}, A. and {Molinaro}, R. and {Muraveva}, T. and {Leccia}, S. and {Rimoldini}, L. and {Holl}, B. and {Jevardat de Fombelle}, G. and {Sartoretti}, P. and {Marchal}, O. and {Audard}, M. and {Nienartowicz}, K. and {Andrae}, R. and {Marconi}, M. and {Szabados}, L. and {Evans}, D.~W. and {Lecoeur-Taibi}, I. and {Mowlavi}, N. and {Musella}, I. and {Eyer}, L.},
        title = "{Gaia Data Release 3. Specific processing and validation of all-sky RR Lyrae and Cepheid stars: The RR Lyrae sample}",
      journal = {\aap},
         year = 2023,
        month = jun,
       volume = {674},
          eid = {A18},
        pages = {A18},
          doi = {10.1051/0004-6361/202243964},
archivePrefix = {arXiv},
       eprint = {2206.06278},
 primaryClass = {astro-ph.SR},
       adsurl = {https://ui.adsabs.harvard.edu/abs/2023A&A...674A..18C}
}

@ARTICLE{Heinze2018,
       author = {{Heinze}, A.~N. and {Tonry}, J.~L. and {Denneau}, L. and {Flewelling}, H. and {Stalder}, B. and {Rest}, A. and {Smith}, K.~W. and {Smartt}, S.~J. and {Weiland}, H.},
        title = "{A First Catalog of Variable Stars Measured by the Asteroid Terrestrial-impact Last Alert System (ATLAS)}",
      journal = {\aj},
         year = 2018,
        month = nov,
       volume = {156},
       number = {5},
          eid = {241},
        pages = {241},
          doi = {10.3847/1538-3881/aae47f},
archivePrefix = {arXiv},
       eprint = {1804.02132},
 primaryClass = {astro-ph.SR},
       adsurl = {https://ui.adsabs.harvard.edu/abs/2018AJ....156..241H}
}

@ARTICLE{Cantat2018,
       author = {{Cantat-Gaudin}, T. and {Vallenari}, A. and {Sordo}, R. and {Pensabene}, F. and {Krone-Martins}, A. and {Moitinho}, A. and {Jordi}, C. and {Casamiquela}, L. and {Balaguer-N{\'u}nez}, L. and {Soubiran}, C. and {Brouillet}, N.},
        title = "{Characterising open clusters in the solar neighbourhood with the Tycho-Gaia Astrometric Solution}",
      journal = {\aap},
         year = 2018,
        month = jul,
       volume = {615},
          eid = {A49},
        pages = {A49},
          doi = {10.1051/0004-6361/201731251},
archivePrefix = {arXiv},
       eprint = {1801.10042},
 primaryClass = {astro-ph.GA},
       adsurl = {https://ui.adsabs.harvard.edu/abs/2018A&A...615A..49C}
}

@ARTICLE{Drake2014,
       author = {{Drake}, A.~J. and {Graham}, M.~J. and {Djorgovski}, S.~G. and {Catelan}, M. and {Mahabal}, A.~A. and {Torrealba}, G. and {Garc{\'\i}a-{\'A}lvarez}, D. and {Donalek}, C. and {Prieto}, J.~L. and {Williams}, R. and {Larson}, S. and {Christen sen}, E. and {Belokurov}, V. and {Koposov}, S.~E. and {Beshore}, E. and {Boattini}, A. and {Gibbs}, A. and {Hill}, R. and {Kowalski}, R. and {Johnson}, J. and {Shelly}, F.},
        title = "{The Catalina Surveys Periodic Variable Star Catalog}",
      journal = {\apjs},
         year = 2014,
        month = jul,
       volume = {213},
       number = {1},
          eid = {9},
        pages = {9},
          doi = {10.1088/0067-0049/213/1/9},
archivePrefix = {arXiv},
       eprint = {1405.4290},
 primaryClass = {astro-ph.SR},
       adsurl = {https://ui.adsabs.harvard.edu/abs/2014ApJS..213....9D}
}

@ARTICLE{Murphy2019,
       author = {{Murphy}, Simon J. and {Hey}, Daniel and {Van Reeth}, Timothy and {Bedding}, Timothy R.},
        title = "{Gaia-derived luminosities of Kepler A/F stars and the pulsator fraction across the {\ensuremath{\delta}} Scuti instability strip}",
      journal = {\mnras},
         year = 2019,
        month = may,
       volume = {485},
       number = {2},
        pages = {2380-2400},
          doi = {10.1093/mnras/stz590},
archivePrefix = {arXiv},
       eprint = {1903.00015},
 primaryClass = {astro-ph.SR},
       adsurl = {https://ui.adsabs.harvard.edu/abs/2019MNRAS.485.2380M}
}

@ARTICLE{2011AJ....142..110M,
       author = {{McNamara}, D.~H.},
        title = "{Delta Scuti, SX Phoenicis, and RR Lyrae Stars in Galaxies and Globular Clusters}",
      journal = {\aj},
         year = 2011,
        month = oct,
       volume = {142},
       number = {4},
          eid = {110},
        pages = {110},
          doi = {10.1088/0004-6256/142/4/110},
       adsurl = {https://ui.adsabs.harvard.edu/abs/2011AJ....142..110M}
}

@ARTICLE{2019MNRAS.486.4348Z,
       author = {{Ziaali}, Elham and {Bedding}, Timothy R. and {Murphy}, Simon J. and {Van Reeth}, Timothy and {Hey}, Daniel R.},
        title = "{The period-luminosity relation for {\ensuremath{\delta}} Scuti stars using Gaia DR2 parallaxes}",
      journal = {\mnras},
         year = 2019,
        month = jul,
       volume = {486},
       number = {3},
        pages = {4348-4353},
          doi = {10.1093/mnras/stz1110},
archivePrefix = {arXiv},
       eprint = {1904.08101},
 primaryClass = {astro-ph.SR},
       adsurl = {https://ui.adsabs.harvard.edu/abs/2019MNRAS.486.4348Z}
}

@ARTICLE{Stassun2019,
       author = {{Stassun}, Keivan G. and {Oelkers}, Ryan J. and {Paegert}, Martin and {Torres}, Guillermo and {Pepper}, Joshua and {De Lee}, Nathan and {Collins}, Kevin and {Latham}, David W. and {Muirhead}, Philip S. and {Chittidi}, Jay and {Rojas-Ayala}, B{\'a}rbara and {Fleming}, Scott W. and {Rose}, Mark E. and {Tenenbaum}, Peter and {Ting}, Eric B. and {Kane}, Stephen R. and {Barclay}, Thomas and {Bean}, Jacob L. and {Brassuer}, C.~E. and {Charbonneau}, David and {Ge}, Jian and {Lissauer}, Jack J. and {Mann}, Andrew W. and {McLean}, Brian and {Mullally}, Susan and {Narita}, Norio and {Plavchan}, Peter and {Ricker}, George R. and {Sasselov}, Dimitar and {Seager}, S. and {Sharma}, Sanjib and {Shiao}, Bernie and {Sozzetti}, Alessandro and {Stello}, Dennis and {Vanderspek}, Roland and {Wallace}, Geoff and {Winn}, Joshua N.},
        title = "{The Revised TESS Input Catalog and Candidate Target List}",
      journal = {\aj},
         year = 2019,
        month = oct,
       volume = {158},
       number = {4},
          eid = {138},
        pages = {138},
          doi = {10.3847/1538-3881/ab3467},
archivePrefix = {arXiv},
       eprint = {1905.10694},
 primaryClass = {astro-ph.SR},
       adsurl = {https://ui.adsabs.harvard.edu/abs/2019AJ....158..138S}
}

@ARTICLE{Bowman2018,
       author = {{Bowman}, Dominic M. and {Kurtz}, Donald W.},
        title = "{Characterizing the observational properties of {\ensuremath{\delta}} Sct stars in the era of space photometry from the Kepler mission}",
      journal = {\mnras},
         year = 2018,
        month = may,
       volume = {476},
       number = {3},
        pages = {3169-3184},
          doi = {10.1093/mnras/sty449},
archivePrefix = {arXiv},
       eprint = {1802.05433},
 primaryClass = {astro-ph.SR},
       adsurl = {https://ui.adsabs.harvard.edu/abs/2018MNRAS.476.3169B}
}

@ARTICLE{Hasanzadeh2021,
       author = {{Hasanzadeh}, A. and {Safari}, H. and {Ghasemi}, H.},
        title = "{Relations between the asteroseismic indices and stellar parameters of {\ensuremath{\delta}} Scuti stars for two years of TESS mission}",
      journal = {\mnras},
         year = 2021,
        month = jul,
       volume = {505},
       number = {1},
        pages = {1476-1484},
          doi = {10.1093/mnras/stab1411},
       adsurl = {https://ui.adsabs.harvard.edu/abs/2021MNRAS.505.1476H}
}

@ARTICLE{Petersen1996,
       author = {{Petersen}, J.~O. and {Christensen-Dalsgaard}, J.},
        title = "{Pulsation models of {\ensuremath{\delta}} Scuti variables. I. The high-amplitude double-mode stars.}",
      journal = {\aap},
         year = 1996,
        month = aug,
       volume = {312},
        pages = {463-474},
       adsurl = {https://ui.adsabs.harvard.edu/abs/1996A&A...312..463P}
}

@ARTICLE{Poretti2005,
       author = {{Poretti}, E. and {Su{\'a}rez}, J.~C. and {Niarchos}, P.~G. and {Gazeas}, K.~D. and {Manimanis}, V.~N. and {van Cauteren}, P. and {Lampens}, P. and {Wils}, P. and {Alonso}, R. and {Amado}, P.~J. and {Belmonte}, J.~A. and {Butterworth}, N.~D. and {Martignoni}, M. and {Mart{\'\i}n-Ruiz}, S. and {Moskalik}, P. and {Robertson}, C.~W.},
        title = "{The double-mode nature of the HADS star GSC 00144-03031 and the Petersen diagram of the class}",
      journal = {\aap},
         year = 2005,
        month = sep,
       volume = {440},
       number = {3},
        pages = {1097-1104},
          doi = {10.1051/0004-6361:20053463},
archivePrefix = {arXiv},
       eprint = {astro-ph/0506266},
 primaryClass = {astro-ph},
       adsurl = {https://ui.adsabs.harvard.edu/abs/2005A&A...440.1097P}
}

@ARTICLE{Furgoni2016,
       author = {{Furgoni}, R.},
        title = "{Analysis of the Petersen Diagram of Double Mode High Amplitude delta Scuti Stars}",
      journal = {\jaavso},
         year = 2016,
        month = jun,
       volume = {44},
       number = {1},
        pages = {6},
          doi = {10.48550/arXiv.1602.07254},
archivePrefix = {arXiv},
       eprint = {1602.07254},
 primaryClass = {astro-ph.SR},
       adsurl = {https://ui.adsabs.harvard.edu/abs/2016JAVSO..44....6F}
}

@ARTICLE{Yang2021,
       author = {{Yang}, Tao-Zhi and {Zuo}, Zhao-Yu and {Wang}, Xin-Yue and {Sun}, Xiao-Ya and {Tang}, Rui-Xuan},
        title = "{A catalogue of double-mode high amplitude $\delta$ Scuti stars in the Galaxy and their statistical properties}",
      journal = {arXiv e-prints},
         year = 2021,
        month = oct,
          eid = {arXiv:2110.13594},
        pages = {arXiv:2110.13594},
          doi = {10.48550/arXiv.2110.13594},
archivePrefix = {arXiv},
       eprint = {2110.13594},
 primaryClass = {astro-ph.SR},
       adsurl = {https://ui.adsabs.harvard.edu/abs/2021arXiv211013594Y}
}

@ARTICLE{Foreman-Mackey2013,
       author = {{Foreman-Mackey}, Daniel and {Hogg}, David W. and {Lang}, Dustin and {Goodman}, Jonathan},
        title = "{emcee: The MCMC Hammer}",
      journal = {\pasp},
         year = 2013,
        month = mar,
       volume = {125},
       number = {925},
        pages = {306},
          doi = {10.1086/670067},
archivePrefix = {arXiv},
       eprint = {1202.3665},
 primaryClass = {astro-ph.IM},
       adsurl = {https://ui.adsabs.harvard.edu/abs/2013PASP..125..306F}
}

@ARTICLE{Dziembowski1977,
       author = {{Dziembowski}, W.},
        title = "{Oscillations of giants and supergiants.}",
      journal = {\actaa},
         year = 1977,
        month = jan,
       volume = {27},
        pages = {95-126},
       adsurl = {https://ui.adsabs.harvard.edu/abs/1977AcA....27...95D}
}

@ARTICLE{suarez2006,
       author = {{Su{\'a}rez}, J.~C. and {Garrido}, R. and {Goupil}, M.~J.},
        title = "{The role of rotation on Petersen diagrams. The {\ensuremath{\Pi}}$_{1/0 ({\ensuremath{\Omega}})}$ period ratios}",
      journal = {\aap},
         year = 2006,
        month = feb,
       volume = {447},
       number = {2},
        pages = {649-653},
          doi = {10.1051/0004-6361:20053866},
archivePrefix = {arXiv},
       eprint = {astro-ph/0510128},
 primaryClass = {astro-ph},
       adsurl = {https://ui.adsabs.harvard.edu/abs/2006A&A...447..649S}
}

@ARTICLE{suarez2007,
       author = {{Su{\'a}rez}, J.~C. and {Garrido}, R. and {Moya}, A.},
        title = "{The role of rotation on Petersen diagrams. II. The influence of near-degeneracy}",
      journal = {\aap},
         year = 2007,
        month = nov,
       volume = {474},
       number = {3},
        pages = {961-967},
          doi = {10.1051/0004-6361:20077647},
archivePrefix = {arXiv},
       eprint = {0709.2006},
 primaryClass = {astro-ph},
       adsurl = {https://ui.adsabs.harvard.edu/abs/2007A&A...474..961S}
}

\end{document}